\documentclass[preprint2]{aastex}

\slugcomment{Not to appear in Nonlearned J., 45.}

\shorttitle{observational study on the chromospheric evaporation of an X 1.6 flare}
\shortauthors{Lee et al.}

\begin{document}

\title{{\it IRIS}, {\it Hinode}, {\it SDO}, and {\it RHESSI} observations of a white light flare produced directly by non-thermal electrons}

\author{Kyoung-Sun Lee}
\affil{Hinode Science Center, National Astronomical Observatory of Japan (NAOJ), 2-21-1, Osawa, Mitaka, Tokyo 181-8588, Japan}

\email{ksun.lee@nao.ac.jp}

\author{Shinsuke Imada}
\affil{Institute for Space-Earth Environmental Research (ISEE), Nagoya University, Furo-cho, Chikusa-ku, Nagoya 466-8550, Japan}

\author{Kyoko Watanabe}
\affil{National Defense Academy of Japan, 1-10-20 Hashirimizu, Yokosuka 239-8686, Japan}

\author{Yumi Bamba}
\affil{Hinode team, ISAS/JAXA, 3-1-1 Yoshinodai, Chuo-ku, Sagamihara, Kanagawa 252-5210, Japan}

\and
\author{David H. Brooks \altaffilmark{1}}
\affil{College of Science, George Mason University, 4400 University Drive, Fairfax, VA 22030, USA}
\altaffiltext{1}{Current address: Hinode Team, ISAS/JAXA, 3-1-1 Yoshinodai, Chuo-ku, Sagamihara, Kanagawa 252-5210, Japan}

\begin{abstract} 
 An X1.6 flare occurred in AR 12192 on 2014 October 22 at 14:02 UT and was observed by {\it Hinode}, {\it IRIS}, {\it SDO}, and {\it RHESSI}. We analyze a bright kernel which produces a white light (WL) flare with continuum enhancement and a hard X-ray (HXR) peak. Taking advantage of the spectroscopic observations of {\it IRIS} and {\it Hinode}/EIS, we measure the temporal variation of the plasma properties in the bright kernel in the chromosphere and corona. We found that explosive evaporation was observed when the WL emission occurred, even though the intensity enhancement in hotter lines is quite weak. The temporal correlation of the WL emission, HXR peak, and evaporation flows indicate that the WL emission was produced by accelerated electrons. 
To understand the white light emission process, we calculated the energy flux deposited by non-thermal electrons (observed by {\it RHESSI}) and compared it to the dissipated energy estimated from a 
chromospheric line (\ion{Mg}{2} triplet) observed by {\it IRIS}. The deposited energy flux from the non-thermal electrons is about $3 \sim 7.7 \times 10^{10} \rm{erg ~ cm^{-2} ~ s^{-1}}$ for a given low energy cut-off of $30 \sim 40\, {\rm keV}$, assuming the thick target model. The energy flux estimated from the temperature changes in the chromosphere measured using the \ion{Mg}{2} subordinate line is about $4.6-6.7 \times 10^{9} \rm {erg ~ cm^{-2} ~ s^{-1}}$: $\sim 6-22\%$ of the deposited energy. 
This comparison of estimated energy fluxes implies that the continuum enhancement was directly produced by the non-thermal electrons.

 \end{abstract}

 \keywords{Sun: activity --- Sun: chromosphere --- Sun: corona --- Sun: flares --- Sun: UV radiation --- techniques: spectroscopic}

 \section{Introduction}
Solar flares are one of the most energetic energy release processes in the heliosphere. When a flare occurs we can observe the multi-wavelength response from microwaves to X-rays such as radio bursts in the impulsive phase, H$\alpha$ emission along the flare ribbons, soft X-ray emission in the post flare loop arcade, and hard X-ray emission at the footpoints and looptop region \citep{fletcher_etal2011}. 

Based on these multi-wavelength observations, several solar flare models have been proposed. The standard solar flare model (CSHKP; \citet{carmichael_1964, sturrock_1968, hirayama_1974, kopp&pneuman_1976}) proposes that magnetic reconnection occurs at coronal heights and the released magnetic energy is transported to the lower atmospheric layers (e.g. the chromosphere) by thermal conduction \citep{nagai_1980, yokoyama&shibata_2001}, non-thermal particles \citep{nagai&emslie_1984, fisher_etal1985a, fisher_etal1985b, fisher_etal1985c}, or an Alfv\'en wave poynting flux \citep{fletcher_hudson2008}. The transferred energy heats the plasma and generates and overpressure in the lower atmosphere. Dense plasma is then evaporated toward the corona along the magnetic field (chromospheric evaporation) and we then observe post flare loops emitting in the EUV and soft X-rays.

Sometimes strong flares produce continuum enhancements as a photospheric response during the impulsive phase of the flare, and this is termed a white light flare (WLF) \citep{carrington_1859, svestka_1966}. Previous observations in visible wavelengths and hard X-rays showed that the continuum enhancement in WLFs is well correlated with hard X-ray emission both spatially and temporally \citep{neidig_1989, hudson_etal1992, metcalf_etal2003, watanabe_etal2010, krucker_etal2015, kuhar_etal2016}. As a result of this correlation, it has been thought that WLFs are produced by the transported energy from accelerated particles such as non-thermal electrons \citep{brown_1971, hudson_1972}. 

With recent high spatial resolution observations, white light emission has been reported even in C-class flares \citep{matthews_etal2003, hudson_etal2006}. The electron flux from these low energy events is not enough to penetrate and heat the photosphere directly. Therefore, other heating mechanisms have also been considered. For example, \citet{zirin&neidig_1981} proposed that high energy protons carry the energy, and \citet{machado_etal1989} suggested that WL emission is produced by electrons that heat the chromosphere directly and the photosphere indirectly through radiative backwarming. However, the true heating mechanism in the lower atmosphere remains unclear. We believe that understanding the flare dynamics during the impulsive phase, and the lower atmospheric response to the flare, are key to clarifying the heating and energy transport processes.

From previous observational studies of chromospheric evaporation,  
strong blue shifted emission ($>100~\rm{km~ s^{-1}}$) in coronal lines \citep{antonucci_etal1982, brosius2003, brosius2009, brosius2013a, brosius2013b, milligan&dennis_2009} and red asymmetries  ($40-100~\rm{km~ s^{-1}}$) in chromospheric lines \citep{ichimoto&kurokawa_1984, kamio_etal2005, delzanna_etal2006} have been found. 
Some flares also show a redshift in coronal lines formed around a few MK \citep{imada_etal2008, milligan&dennis_2009, watanabe_t_etal2010}, and direct imaging observations of chromospheric evaporation upflows have been observed by Hinode/XRT \citep{nitta_etal2012}. Hydrodynamic simulations can re-produce aspects of the observations and predict two different types of evaporative flows depending on the deposited energy flux: "explosive" and "gentle" evaporation \citep{fisher_etal1985c, fisher_etal1985b, fisher_etal1985a}. 

Recently, the {\it Hinode}/EUV Imaging Spectrometer (EIS, \citet{culhane_etal2007a}) and {\it Interface Region Imaging Spectrograph} ({\it IRIS}, \citet{depontieu_etal2014}) have provided us with high spatial and temporal resolution spectroscopic observations in the EUV/UV \citep{li&ding_2011,tian_etal2015,graham&cauzzi_2015}. The combined power of the instruments allows us to investigate flare properties and dynamics through the entire atmospshere from chromosphere to corona \citep{li_y_etal2015, polito_etal2015, polito_etal2016}.

Most relevant to this work are several studies comparing the deposited flare energy with observed continuum enhancements.
For example, \citet{watanabe_etal2010} 
found that the energy of the white light emission observed by {\it Hinode}/Solar Optical Telescope (SOT) was equivalent to the energy supplied by all the electrons accelerated to above 40 keV, which suggests that highly accelerated electrons are responsible for producing white light emission. 
Recently, \citet{kleint_etal2016} investigated the radiated energy from the continuum enhancement observed from the UV to IR during a flare using {\it IRIS}, {\it Solar Dynamics Observatory} ({\it SDO}), and {\it Facility Infrared Spectrometer} ({\it FIRS}), and also found that 
the deposited energy was sufficient to produce the UV and visible continuum emission in the flare. 

In this study, we describe the temporal evolution of the spectral properties and quantitatively estimate the energy flux of an X1.6 flare using combined observations from {\it IRIS}, EIS, {\it Reuven Ramaty High-Energy Solar Spectroscopic Imager} ({\it RHESSI}, \citet{lin_etal2002}), and {\it SDO}. The whole flare evolution, from the beginning of the impulsive phase to the gradual phase, is captured by these instruments simultaneously and the strong flare produces white light emission. 
Previously, \citet{li_d_etal2015}, \citet{thalmann_etal2015}, and  \citet{veronig&polanec_2015} 
also investigated this well observed flare using SDO, RHESSI and {\it IRIS}.  \citet{thalmann_etal2015} and \citet{veronig&polanec_2015} mainly focused on the magnetic reconnection rates and the RHESSI HXR profiles of this flare, and \citet{li_d_etal2015} investigated the relationship between Doppler velocity patterns derived from IRIS \ion{Fe}{21} and \ion{C}{1} and the RHESSI HXR intensity. Those studies showed that the flare energy is injected into high energy electrons, and that they could driver the evaporation flow in the flare.
We present the flare observations from each instrument in Section 2, and the temporal evolution of the spectral properties (intensity, Doppler velocity, line width, density and temperature) of the flare kernel during the impulsive phase with the other continuum and X-ray observations in Section 3. We discuss the comparison between the deposited energy from the non-thermal electrons and the observed spectroscopic properties in Section 4. A summary of our results is given in Section 5.

\section{Observations and data analysis}

\subsection{X1.6 flare in AR 12192}
We investigate an X1.6 flare which occurred on 2014 October 22 in AR 12192. AR 12192 was the largest active region in this solar cycle and it produced 6 X-class flares and 31 M-class flares. 
Figure 1 shows the {\it GOES} soft X-ray (0.5-4 $\rm{\AA}$ and 1-8 $\rm{\AA}$) light curve, its time derivative, and the {\it RHESSI} hard and soft X-ray light curves of the flare event. 
The flare started at about 14:02 UT and peaked at 14:28 UT. The gradual phase of the flare emission declined until an M-class flare occurred at 15:54 UT. The vertical dashed lines mark the specific times at which we present the spectral properties of the flare:  
(a: the beginning of the impulsive phase (14:06 UT), b: the rise phase (14:09 UT), and c: the peak (14:24 UT); see section 3).
Positions a, b, and c also mark the peaks of the time derivative of the soft X-ray curve which correspond to the hard X-ray peaks (bottom), as expected from the Neupert effect \citep{neupert_1968}.

Figure 2 shows context images of the flare obtained by {\it SDO}/Atmospheric Imaging Assembly (AIA, \citet{lemen_etal2012}) in the 211$~\rm{\AA}$ (a-c) and 1700$~\rm{\AA}$ (e-g) channels, and {\it IRIS} \ion{C}{2} 1330$~\rm{\AA}$ slit jaw images (SJIs) overlaid with EIS 195$~\rm{\AA}$ contours (panels i-k) for different timings, before the flare ($\sim$13:34 UT), the first HXR peak ($\sim$14:06 UT) and the SXR peak ($\sim$14:24 UT). Panels (d) and (h) display the {\it SDO}/Helioseismic and Magnetic Imager (HMI, \citet{schou_etal2012}) continuum and a running difference image at the time the white light flare occurred, respectively. Panel (l) shows the polarity inversion line from the HMI magnetogram contoured on an {\it IRIS} SJI to show the magnetic field configuration. 

When we look at the continuum and lower atmospheric (AIA 1700$~\rm{\AA}$) images, we can see that the flare occurred at the boundary of the large umbra and the satellite penumbra. When the first hard X-ray peak of the flare was observed (around 14:06 UT), we can see continuum enhancements (the white light flare) at two bright kernels (panel (h)) which correspond to the footpoints of the flare loop structure (panel (c)). Then, two ribbons extend in the east-west direction (panels (g) and (k)). Before the flare, around 13:34 UT (panels (a), (e), and (i)), we can see a small brigtening near the east side of the flare kernel. This kernel was observed simultaneously in the {\it IRIS} and EIS scanning rasters. We analyzed the evolution of the plasma properties of this bright kernel and estimated the energy flux.

\subsection{Spectroscopic observations from {\it Hinode}/EIS and {\it IRIS}}
EIS captured the flare with the study ``HH\_Flare\_raster\_v6" which ran from 13:01:56 UT to 15:56:56 UT. This EIS study is designed for observing flares using a moderate cadence raster scan. The 2$\arcsec$ slit scans 20 positions with a coarse 3$\arcsec$ step between each position and the field of view is 59$\arcsec \times 152\arcsec$. The exposure time at each position is 9 seconds and the raster scan takes about 3.5 minutes. The study has 12 spectral windows and we used 10 spectral lines covering the temperature range from log T = 4.9-7.2. These are listed in Table 1. The spectral resolution of the EIS is about 0.022$~\rm{\AA}$. 

At the same time, {\it IRIS} was running a very large coarse 8 step raster. It uses the 0.33$\arcsec ~\times $ 175$\arcsec$ slit and with 2$\arcsec$ steps and so covers a field of view of about 14$\arcsec ~\times $ 175$\arcsec$ in around 130 seconds. The exposure time at each position is 16 seconds. The spectral and spatial resolution of {\it IRIS} is 0.025 $\rm{\AA}$ and 0.32$\arcsec$, respectively. The {\it IRIS} slit direction was rotated 45 degrees relative to its center for this observation. The observing program includes 9 spectral windows in the FUV (1332-1358$~\rm{\AA}$ and 1389-1407$~\rm{\AA}$) and NUV (2783-2834$~\rm{\AA}$). In this study, we only analyze the spectral lines which are close to optically thin, \ion{O}{1}, \ion{Si}{4}, and \ion{Fe}{21}, for measuring the Doppler velocity, and used the \ion{O}{4} and \ion{Mg}{2} lines for investigating the chromospheric response. The spectral lines we used are summarized in Table 1.

To obtain the intensity, Doppler velocity, and line width as a function of time, we fitted the spectral lines in Table 1 using single and multiple gaussians. 
Figures 3 and 4 show examples of the line profiles from EIS and {\it IRIS}
, respectively. 
In Figure 3, the green solid line is the fitted spectra and the red dashed lines indicate each component of the multiple Gaussian fitting. The red and green dotted vertical lines are the fitted line center of each spectral line component from a multiple Gaussian fitting, and the estimated velocities are written in the Figure 3. To obtain a reference wavelength for the EIS spectra, we measured the average line centers before the flare (between 13:01-13:51 UT), and these are marked with vertical dashed lines. In Figure 4, the green solid line is the fitted spectra and the dashed lines indicate each component of the multiple Gaussian fits. The red dashed fitted components are \ion{Fe}{21} (in the upper panels) and the second component of \ion{Si}{4} (in the lower panels). The reference wavelengths for {\it IRIS} were determined by taking the difference between the theoretical wavelengths and averaged observed wavelengths of O I 1355.60$~\rm{\AA}$ and S I 1401.51$~\rm{\AA}$ before the flare. These are marked by vertical dashed lines in the Figure. The theoretical wavelength of \ion{Fe}{21} is taken from the {\it CHIANTI} atomic database \citep{dere_etal1997, landi_etal2012} and is denoted by the vertical dot-dashed line. 

\subsection{{\it SDO} and {\it RHESSI} observations}
We also used AIA and HMI onboard {\it SDO} to understand the global structure and magnetic field configuration of the flare. AIA provides multiple temperature images covering log T = 3.7-7.2 with a high time cadence of about 12 seconds and a spatial resolution of 1.2 arcsec. We used AIA 1700 $\rm{\AA}$ and 211 $\rm{\AA}$ filter images for context (Figure 2), and took advantage of the high temporal resolution of AIA to investigate the flare intensity variations and compare them to what is observed by EIS and {\it IRIS}. The temporal variation of the intensity in different AIA filters is plotted in Figure 5. HMI provides full Sun line of sight (LOS) magnetograms and continuum images at a spatial resolution of $\sim 1 \arcsec$ and a temporal cadence of 45 seconds. Using the continuum data, we confirmed that the white light flare kernel is coincident with the same bright UV kernel we analyzed. 

We also investigated the flare hard and soft X-ray emission using the {\it RHESSI} X-ray spectrometer. We plotted the light curve of the emission in the 30-100 keV and 12-25 keV range in Figure 1. To obtain images of the HXR and SXR emission, we used the "Clean" method with 300 iterations and a temporal resolution of 2 minutes, which is similar to that of the {\it IRIS} raster. The left panel in Figure 6 shows an HMI intensity difference image with the cleaned {\it RHESSI} HXR and SXR intensity contours (50, 60, 70, 80, and 90$\%$) overlaid.


\subsection{Co-alignment of {\it SDO}, {\it Hinode}/EIS, and {\it IRIS}}
We co-aligned the flare observations from {\it Hinode}, {\it IRIS}, {\it RHESSI}, and {\it SDO} as follows. First, we used {\it SDO}/AIA observations as the reference image and aligned the AIA 1600 $\rm{\AA}$ image with the {\it IRIS} SJI 1330 $\rm{\AA}$ images. The {\it IRIS} slit was rotated 45 degrees from the north-south direction, so we de-rotated the SJI images and then aligned them with the AIA images. Then, we aligned the EIS \ion{Fe}{12} 195.12 $\rm{\AA}$ raster images with AIA 193 $\rm{\AA}$ filter images. For this alignment we calculated the offset values using the procedure `$align\_map.pro$' available in the Solar Software (SSW) package. The offset values vary within $\sim$2\arcsec, and the EIS and {\it IRIS} SJI images are overlaid in the bottom panels of Figure 2.

\section{Results}

\subsection{Temporal evolution of the continuum, UV, and X-ray emission in the bright kernel}
We found that the white light flare signature of the bright kernel in the HMI images and the HXR emission observed by {\it RHESSI} were located at the same position (left panel in Figure 6). To check the temporal evolution we plotted the light curve of the HMI continuum intensity in the bottom panel of Figure 7, and the HXR and SXR light curves for the bright kernel location in the first and second rows. We also plotted the light curves for the chromospheric and flaring line intensities from {\it IRIS} (\ion{O}{1}, \ion{Si}{4}, \ion{Fe}{21}) for comparison. 

First, we note that the HXR peak and enhancement of the chromospheric line intensity (\ion{O}{1}) appear at the beginning of the flare ($\sim$14:04 UT). Second, the continuum enhancement starts to appear around the same time and reaches its maximum within 2 minutes. 
Third, the SXR emission shows two borad humps, one near the HXR peak time and another lower peak around 16 minutes later. This suggests that the heated plasma was evaporated and the increased density and temperature are observed as enhanced SXR emission. Fourth, the emission from \ion{Fe}{21} increases around 14:24 UT.
We note that even although the intensity of \ion{Fe}{21} is weak, emission is detected at the beginning of the flare (14:04 $\sim$ 14:11 UT). It suggests that the enhanced \ion{Fe}{21} emission might be caused by the density enhancement of the hot plasma. 

\subsection{Temporal evolution of the spectral properties of the bright kernel}


  
 \subsubsection{Intensity}
Figures 5, 8, and 10 (upper panel) show the temporal variation of the intensities of the bright kernel at different wavelengths from {\it SDO}, EIS, and {\it IRIS}, respectively. The {\it SDO}/AIA intensities in Figure 5 are normalized by their maximum intensity during the period from 13:00 UT to 16:00 UT. 
Together EIS and {\it IRIS} provide the temporal variation of the intensity over a wide range of temperature from log T=4.5 - 7.2. 

From the intensity variation, we can see analogous behavior in certain temperature ranges, log T= 4.5 - 5.8 (cooler), log T= 5.8 - 6.4 (middle), and log T= 6.4 - 7.2 (hotter). It seems that the temperature response of the flare is similar in these temperature ranges and transitions at log T$\sim$ 5.8  and log T$\sim$ 6.4. In the cooler temperature emission, such as 1600 $\rm{\AA}$ and 1700 $\rm{\AA}$ in Figure 5 or \ion{He}{2} or \ion{O}{5} in Figure 8, the response of the flare is seen as a sharp intensity enhancement when the flare starts (time (a) in Figure 1) with no significant enhancement during the gradual phase. For the hotter emission lines (e.g. \ion{Fe}{24} and \ion{Fe}{23}), it seems that the intensities peak a little later than in the cooler temperature lines and there is a significant intensity enhancement during the gradual phase. In the middle temperature range, e.g. \ion{Fe}{10} - \ion{Fe}{16} in Figure 8, there are several peaks during the gradual phase. It seems that after the main flare, there are still bursts of intensity enhancement albeit not strong enough to produce white light or HXR emission. 

 \subsubsection{Doppler velocity}
The upper panel in Figure 9 and middle row in Figure 10 show the temporal variation of the Doppler velocity at different wavelengths. The most significant velocity variation is seen at the impulsive phase of the flare around 14:06 UT. A strong blue shift is observed in the flaring temperature lines, \ion{Fe}{15} - \ion{Fe}{24}, and a red shift is observed in the chromospheric lines, \ion{O}{5} and \ion{Si}{4}, which is consistent with the Doppler velocity pattern expected from explosive evaporation (Figures 3 and 9). 
After the impulsive phase the blue shift in the higher temperature lines quickly changes to a weak red shift which lasts more than an hour; essentially until the hotter emission disappears (Figure 10). 
A strong red shift is also seen in the cooler temperature lines during the impulsive phase and weak red-shifted emission is observed during the gradual phase. 

We plotted the velocity as a function of the line formation temperature in Figure 11. Panels (a), (b), and (c) show the velocity pattern before the flare, during the impulsive phase, and during the gradual phase. The velocity pattern we discussed is clearly seen in the Figure. Moreover, the velocity become larger at higher temperatures, which is a similar behavior to what has been reported previously by e.g. \citet{milligan&dennis_2009} and \citet{polito_etal2016}, and is consistent with theoretical expectations \citep{nagai&emslie_1984}.  

We note that a strong blue shift was only observed at the beginning of the flare even though the intensity enhancement in the hotter lines is quite weak. Figure 10 shows that the \ion{Fe}{21} intensity from {\it IRIS} is weak, but the Doppler velocity shows a strong outflow at the start time of the flare ($\sim$ 14:06 UT). The EIS observations also show that a strong blue shift is observed in the hotter lines even when the intensity is weak (Figures 3 and 8). So, it appears that higher temperature emission exists and we can observe the dynamics even when there is no strong intensity signature. The intensity may be weak because the density is low in the high temperature plasma in the early phase, which we tried to verify with density diagnostic measurements (see below). 


 \subsubsection{Line width}
 We also checked the line width variation with time, which is shown in the bottom panels of Figures 9 and 10. The strongest line width enhancement was observed in the flaring emission lines (\ion{Fe}{21} and \ion{Fe}{23}) at $\sim$14:06 UT when the first hard X-ray peak and Doppler velocity peak in the hot plasma appears. We note that if the emission has a bulk Doppler shifted velocity component, it will make the line broader due to the combination of the rest and moving component. However, at the beginning of the flare, the Doppler velocity pattern of the lines can be fitted with a single gaussian, and the whole lines are blue or red shifted without a rest component (see e.g. the line profiles of \ion{Fe}{23} and \ion{Fe}{24} in Figure 3 and \ion{Fe}{21} in Figure 4). 
This implies that the enhanced line width at the beginning of the flare is related to non-thermal broadening, not Doppler velocity. 

The non-thermal broadening in UV and X-ray emissions is usually regarded as a manifestation of unresolved mass motions of the plasma, such as, multiple flows, turbulence or waves \citep{alexander&mackinnon_1993, dere&mason_1993, chae_etal1998b, hara_etal2011, kawate&imada_2013}. The significant enhancement of the line width and the HXR peak at the beginning of the flare are temporally well correlated. If the coronal reconnection occurs at the HXR peak timing, or the HXR comes from the accelerated electrons, the non-thermal broadening possibly caused by turbulence motion or waves due to the magnetic reconnection or accelerated electrons.

 There are several other peaks at different times in \ion{He}{2}, \ion{Si}{4} and \ion{Fe}{12} that appear to be related to small chromospheric brightenings; one of which may be a candidate for lower atmospheric reconnection that triggers the flare \citep{bamba_etal_prep}. 
 
 \subsubsection{Density}
 We measured the density of the bright kernel in the chromosphere and corona during the flare. Assuming the plasma is optically thin, thermal, and in collisional ionization equilibrium, we derive the density using the intensity ratio of emission lines from the allowed and forbidden transitions which are sensitive to the density. The diagnostic method is well described by \citet{mariska_1992_b} and \citet{phillips_etal2008_b}. 
 
The good density sensitive line pairs observed in EIS and IRIS are reported in \citet{young_etal2007a} and \citet{young_2015}, respectively. We used the \ion{O}{4} line pair for measuring the density in the transition region plasma, and the \ion{Fe}{14} line pair for the coronal plasma. Figure 12 shows the temporal variation of the density measured by {\it IRIS} (upper panel) and EIS (lower panel). 

The \ion{O}{4} line pair (1399.77 ${\rm{\AA}}$ and 1401.16 ${\rm{\AA}}$) shows density sensitivity in the Log $N_{e}=10-13$ range with an intensity ratio range of 0.17-0.43. We plotted the intensity ratio with time rather than the converted density because the intensity ratio during the impulsive phase of the flare exceeds the maximum value of the theoretical calculation. Therefore, we cannot measure the density during the impulsive phase. Before and after the impulsive phase, the averaged density is about $6.3\times 10^{10} ~ \rm{cm^{-3}}$. The enhanced intensity ratio during the impulsive phase implies densities in excess of $10^{13} ~\rm{cm^{-3}}$, which could be the result of compression from the explosive evaporation. Alternatively it could mean that the plasma is not in ionization equilibrium \citep{kafatos&tucker1972, imada_etal2011, olluri_etal2013, martinez_etal2016} and therefore that the density measurements are invalid during that period. However, the ionization-relaxation time is only about 13 seconds for a plasma with the average density measured before and after the impulsive phase, which is much shorter than the duration that the intensity ratio was enhanced. 

Another possibility is that the measured line intensities are blended with cool lines which was mentioned by \citet{young_2015} and \citet{polito_etal2016}. For example, \ion{Fe}{2} 1399.96 $\rm{\AA}$, \ion{S}{1} 1401.51 $\rm{\AA}$ and unidentified lines at shorter wavelengths are close to \ion{O}{4}. Even if we perform multiple Gaussian fitting to take these blended lines into consideration, the intensity ratio is still enhanced during the impulsive phase. 
 
We measured the coronal density with EIS using the \ion{Fe}{14} 264.79 $\rm{\AA}$ and 274.20 $\rm{\AA}$ line pair. Compared to the results from \ion{O}{4}, \ion{Fe}{14} 
shows that the density is slightly enhanced in the early phase of the flare, then increases significantly in the later phase and peaks at the same time as the SXR. The temporal variation of the density is similar to the intensity variation, suggesting that the low intensities observed when explosive evaporation occurs are due to the low density of the higher temperature plasma. After evaporation, intensity enhancements can also be seen in the higher temperature lines. 

\subsubsection{\ion{Chromospheric temperature: Mg}{2} triplet lines}
 We also checked the response of the chromospheric \ion{Mg}{2} line. Figure 13 shows the \ion{Mg}{2} line profile and the green horizontal lines mark the region we extracted for the spectral profile. The solid line shows the line profile during the impulsive phase ($\sim$14:07) and the dotted line shows the spectrum around 12:00 UT when there is no specific brightening or X-ray response, as a reference. One interesting point to note is that the \ion{Mg}{2} triplet lines emit strongly during the impulsive phase compared to during the non-flaring time.

 \citet{leenaarts_etal2013b} and \citet{pereira_2015} proposed that \ion{Mg}{2} h \& k, and its subordinate lines (a triplet: 2791.60 $\rm{\AA}$. 2798.75 ${\rm{\AA}}$, and 2798.82 ${\rm{\AA}}$) can be used as diagnostic tools of the chromospheric plasma. In particular, \citet{pereira_2015} showed that the \ion{Mg}{2} triplet blends at 2798.75 ${\rm{\AA}}$ and 2798.82 ${\rm{\AA}}$ will be seen in emission when chromospheric heating occurs, and the line core to wing intensity ratio has a linear relationship with the temperature increase. 

 We applied their quantitative method to investigate the temperature changes in this flare kernel. 
 We measured the line core (the average intensity between 2798.66-2798.93 ${\rm{\AA}}$) to wing (taken at the 2799.32 ${\rm{\AA}}$) intensity ratio of the two blended \ion{Mg}{2} triplet lines, 2798.75 ${\rm{\AA}}$ and 2798.82 ${\rm{\AA}}$.
The left column in Figure 14 shows the variation of the intensity ratio with time. 
We converted the intensity ratio to $\Delta T$ from the linear relationship between them derived from the flare simulation conducted by Pereira (private communication). The variation of $\Delta T$ with time is shown in the right column of Figure 14. The estimated temperature changes are about 
3 kK for 
the flare kernel and the temporal variation shows that the temperature suddenly increased at the beginning of the impulsive phase. 

\subsection{Summary of results}
\begin{itemize}
\item The flare kernel is localized during the impulsive phase and the HXR emission, chromospheric intensity, and white light continuum emission in the kernel are spatially and temporally correlated. The bright kernel peaks first in intensity in the white light continuum and \ion{O}{1}, and then HXR and SXR emission peaks are observed consecutively.

\item The Doppler velocity and line width are enhanced during the impulsive phase. The strongest line width enhancement appears during the first HXR peak, which is just after the white light flare (within a minute). This may be a signature of turbulence from reconnection or heating by non-thermal electrons.

\item Chromospheric (\ion{O}{4}) and coronal (\ion{Fe}{16}) density diagnostics show a strong enhancement during the impulsive phase. The temporal variation indicates that the density is first enhanced in the chromosphere and later in the corona. This is consistent with compression of the chromosphere, the Doppler velocity pattern, and the chromospheric evaporation process.

\item The \ion{Mg}{2} subordinate line blend is seen in emission during the pre-flare and impulsive phase brightenings. The enhancement of the \ion{Mg}{2} core to wing intensity ratio implies the existence of steep temperature gradients and heating at the low atmosphere \citep{pereira_2015}. At the time the white light flare occurs, the ratio becomes over 10, suggesting that strong heating occurs in the lower atmosphere. Moreover, the strong enhancement of the ratio is correlated to the hard X-ray peak, which implies that the non-thermally accelerated electron detected as the HXR emission might be directly related to the low atmospheric heating.

\end{itemize}

\section{Discussion}

\subsection{Spectroscopic results related to the evaporation flow and white light flare}

The flare kernel we discuss in this paper produced a white light flare when the first hard X-ray peak appeared.
The correlation of the HXR peak and Doppler velocity variation in this flare has already been reported by \citet{li_d_etal2015} using {\it IRIS} observations, and they suggested that this implies that the flare is electron driven and that energy deposition from non-thermal electrons produces the chromospheric evaporation flows and white light flare.

From electron beam heating simulations \citep{fisher_etal1985a, fisher_etal1985b, fisher_etal1985c}, it is expected that gentle and explosive evaporation should be observed, depending on the injected energy flux, and that it can be detected by examining the velocity variation at different temperatures. When the energy flux injected is high, the simulation shows a blueshift in coronal lines, {\it "chromospheric~ evaporation"}, and a redshift in chromospheric lines, {\it "chromospheric~ condensation"}. This is called explosive evaporation. 
Conversely, if the injected energy is less than a critical value, about $F_{20}=10^{10} ~ \rm{ergs} ~ \rm{cm^{-2}} ~ \rm{s^{-1}}$, gentle evaporation is observed, and most spectral lines are blue shifted.

Figure 11 shows that explosive evaporation occurs at this bright kernel in the impulsive phase.
To confirm the relationship between the deposited energy from the accelerated electrons and the observed Doppler velocity pattern, we compared the deposited energy flux measured by {\it RHESSI} to the critical value of the energy flux in the simulations.

Assuming this white light flare is produced by accelerated non-thermal electrons, we calculated the total power ($P$) in the non-thermal electrons above a given electron energy (low cut-off energy) under the thick target approximation using the equation \citep{hudson_etal1978, watanabe_etal2010},
\begin{equation}
P(\epsilon \geqslant \epsilon_{c}) = 4.3 \times 10^{24} \frac{b(\gamma)}{\gamma-1} A {\epsilon_{c}}^{-(\gamma-1)} ~ \rm{(erg~ s^{-1})}.
\end{equation}
 For this purpose, we fit the RHESSI hard X-ray photon spectrum. The right panel of Figure 6 shows the fitted {\it RHESSI} spectrum when explosive evaporation occurs (14:06 UT). $\epsilon_{c}$ is the low cut-off energy, $\gamma$ is the spectral index, and $b(\gamma)$ is the auxiliary function from \citet{brown_1971} for the relevant range of $\gamma$.
To measure the energy flux, we determined the size of the hard X-ray emitting region where the HXR (30-100 keV) integrated intensity is greater than 60 $\%$ of the maximum intensity.
The calculated energy fluxes in the non-thermal electrons at the HXR peaks during the impulsive phase ($\sim$14:05 and 14:11 UT), assuming a cut-off energy of 30 keV, is about $7.7 \times 10^{10}~\rm{ergs} ~ \rm{cm^{-2}} ~ \rm{s^{-1}}$ and $6.1 \times 10^{10} ~\rm{ergs} ~ \rm{cm^{-2}} ~ \rm{s^{-1}}$, respectively (Figure 15). The energy of 30 keV is the lowest energy that contains a negligible
amount of thermal emission, and still contains large fluxes of non-thermal photons. We also estimated the deposited energy fluxes by non-thermal HXR electrons above different threshold energies because we don't know which energy electrons affected to the WL emission. For example, non-thermal electrons in low energies cannot penetrate to the photosphere which produces WL emssion. On the other hand, the higher energy electrons may not transport enough energy due to the low photon flux.
Even though we assume the different threshold energies of 40 keV and 50 keV, the energy fluxes at the impulsive phase are about $3.0 \times 10^{10}~\rm{ergs} ~ \rm{cm^{-2}} ~ \rm{s^{-1}}$ and $1.4 \times 10^{10} ~\rm{ergs} ~ \rm{cm^{-2}} ~ \rm{s^{-1}}$, respectively. This shows that the deposited energy from the hard X-rays is strong enough to produce explosive evaporation and is consistent with the observed Doppler velocity pattern.

Furthermore, the flows from explosive evaporation are expected to have a reversal, where the flows change from downflow in the chromosphere to upflow in the corona, and our obsrvations show such a reversal in the temperature range 0.5-2 MK. The velocities in this range also appear to be fairly steady. 
Interestingly, Some downflows, of $5~\rm{km~ s^{-1}}$ in the single gaussian fits and $15~\rm{km~ s^{-1}}$ when a double gaussian is used, are observed in the coronal lines (T$\sim$2MK), which is a surprisingly high temperature. 
A similarly high temperature velocity reversal has previously been reported, however \citep{li&ding_2011}. 

To explain the downflows in the coronal lines, \citet{imada_etal2015} investigated the dependence of the flow reversal temperature on different thermal conduction coefficients. It turns out that if the thermal conduction is strong, the energy is transported quickly by thermal conduction and this scenario shows similar characteristics to the electron beam driven case. If the thermal conduction is weak, however, the energy is mainly transported by the entalphy flux and advection. In this case, the enthalpy flux dominant case, the flow reversal temperature is much hotter than in the thermal conduction case. In our observations, a strong red shift is seen at temperatures around 0.4MK, and the Doppler velocities in the 0.5-2 MK temperature range are mostly steady. This implies that the flow reversal temperature in this flare is not very high, and is similar to that expected in the electron beam model. However, the Doppler velocity around 2MK also shows small downflows, suggesting that we cannot neglect the possibility of energy transport by direct enthalpy flux, in addition to thermal conduction.

At the same time as the Doppler velocity peaks, enhanced line widths are observed at flaring temperatures. This may indicate the presence of turbulence from the non-thermal electrons or reconnection process high in the corona since there is also a temporal correlation between the HXR emission and Doppler velocity peak. Most of the observations show that the electron beam heating model well explains this flare, and the accelerated electrons can produce the white light continuum emission. However, we cannot rule out the possibility of Alfv\'en waves as a heating mechanism \citep{fletcher_hudson2008} given the large non-thermal width in the lower temperature lines, \ion{Fe}{12}, \ion{He}{2}. Recently, \citet{reep&russell_2016} showed that Alfv\'en wave dissipation produces similar heating signatures to electron beam heating, for example, explosive evaporation and a significant temperature enhancement in the chromosphere. If Alfv\'en waves transport the energy flux to the low atmosphere, line widths can also be enhanced by the waves.

\subsection{{\it RHESSI} and {\it IRIS} energy flux comparison}
One of the important issues for understanding the white light flare mechanism is how the energy is transferred to the lower atmosphere to produce the photospheric emission. Recently, it has been reported that not only strong flares but also weak flares (C class flares) can produce white light emission \citep{matthews_etal2003, hudson_etal2006}. So it is important to know how much energy is dissipated in the chromosphere and whether the energy transferred by the electrons is enough to produce a white light flare. Until recently, it has been difficult to estimate the energy flux in the chromosphere due to a lack of observations. After the launch of {\it IRIS}, however, high resolution spectroscopic observations of the chromosphere have become routine, and \citet{pereira_2015} suggested that the \ion{Mg}{2} triplet could be used as a diagnostic tool for quantitatively measuring temperature changes in the chromosphere. 

We have measured the temperature changes during the flare using the \ion{Mg}{2} triplet intensity ratio (section 3.1.5), and used the results to estimate the energy flux deposited in the chromosphere in response to the flare. 
The \ion{Mg}{2} h \& k components show similar peak intensities during the flare implying that they might be in emission even though the plasma is optically thick. Furthermore, the densities measured using the \ion{O}{4} chromospheric line are strongly enhanced during the impulsive phase (Figure 12), which suggests the plasma may be in local thermodynamic equilibrium (LTE). If we assume the plasma is optically thick and in LTE, the energy flux can be determined by Stefan-Boltzmann's law, $F=\sigma {T_{eff}}^{4}$. We estimate the temperature enhancement during the impulsive phase to be $3 \sim 3.3$ kK (Figure 14), so the corresponding energy flux is $4.6 - 6.7 \times 10^{9} ~\rm{ergs~cm^{-2}~s^{-1}}$. Taking this energy flux as an estimate of the amount of energy dissipated in the chromosphere, it is about $ 6 \sim 22~ \%$ of the deposited energy from the accelerated non-thermal electrons measured by {\it RHESSI} assuming the cut-off energy of $ 30 \sim 40 ~\rm{keV}$. This result implies that the majority of the energy from the non-thermal electrons accelerated in the corona is still available to directly produce a white light flare in this event.

Recently, \cite{milligan_etal2014} and \citet{kleint_etal2016} investigated the continuum enhancement across the EUV, UV, visible, and infra-red during a flare, and compared it to the energy deposited by non-thermal electrons observed by {\it RHESSI}. In their investigation, \citet{milligan_etal2014} showed that 15 $\%$ of the deposited energy is radiated by line and continuum emission in the lower atmosphere and \citet{kleint_etal2016} found that 23$\%$ of the deposited energy is radiated by continuum emission. From their investigations, more than 60 $\%$ of the energy is unaccounted for, and they suggested that it is dissipated by heating, plasma motions, or radiated away in other spectral regions or lines. Using {\it IRIS} and HMI continuum observations, we also measured the energy flux from the UV and WL continuum. We converted the observed \ion{Mg}{2} DN to intensity using the {\it iris\_get\_response.pro} routine in SolarSoft and we calibrated the HMI intensity by comparison with the disk intensity reported in the atlas of Brault \& Neckel. We then estimated the radiated power, $P_{\lambda} = \pi ~ I_{\lambda} ~ A ~ \Delta \lambda$, using the intensities ($I_{\lambda}$) and band widths ($\Delta \lambda$) of the IRIS and HMI continua. The energy fluxes from the {\it RHESSI} HXR, {\it IRIS} \ion{Mg}{2} triplet, and HMI continuum are shown in Figure 15 and the energy fluxes in the UV and HMI continuum are only very small fractions of the power in the HXR emission. The estimated energy flux from dissipation by heating in the chromosphere may amount to $\sim 10 \%$ of the deposited energy. 

In this study, even though we have not performed any numerical modeling, we have been able to measure the temperature changes in the chromosphere quantitatively using the \ion{Mg}{2} triplet observed by {\it IRIS}, as suggested by \citet{pereira_2015}. It is the first attempt to apply this diagnostic technique to flare observations, and the results show that the temperature changes in the chromosphere are about 3000 K, which is consistent with the results from the numerical modeling of the UV to IR spectra of \citet{kleint_etal2016}. The results support their suggestion from numerical modeling that quite a strong temperature enhancement is needed in the chromosphere to produce noticeable continuum enhancement in the flare.

\section{Summary}
A bright kernel in an X1.6 flare on 2014 October 22 was observed by {\it Hinode}, {\it IRIS}, {\it SDO}, and {\it RHESSI}. The simultaneous observations covered the whole duration of the flare and the bright kernel produces a multi-wavelength intensity enhancement from continuum to hard X-rays.  We investigated the temporal variation of the spectral properties of this kernel and estimated the energy flux at different wavelengths. 


The multi-wavelength spectroscopic observations show that the flare kernel is localized during the impulsive phase and the HXR emission, chromospheric intensity, and white light continuum emission in the kernel are spatially and temporally correlated. We found that explosive evaporation occurs and there are strong line width enhancements during the first peak of the hard X-ray emission, which is also coincident with the timing of the white light flare. This may indicate that electron beam heating produces strong evaporation flows and there is turbulence from the reconnection or non-thermal electron heating. Furthermore, the \ion{Mg}{2} subordinate line blend is in emission during the impulsive phase. The strong enhancement of the \ion{Mg}{2} core to wing intensity ratio line is correlated with the hard X-ray peak, imply the existence of a steep temperature gradient and heating. The correlated temporal variation of the hard X-ray, white light, explosive evaporation flows, the \ion{Mg}{2} line response and the comparison of the energy flux through the corona to the chromosphere implies that the flare heating and evaporation flow is driven by non-thermal electrons, though we cannot rule out a possible contribution from Alf\'en wave heating.

We have also estimated the energy flux in the bright kernel using the intensities and temperature changes in the chromosphere. Comparison of the deposited energy in the corona and the dissipated energy in the chromosphere shows that accelerated electrons dissipate only $\sim$20 $\%$ of their energy in the chromosphere and the remaining energy is enough to directly produce white light flare emission in the photosphere. The flare we investigated is quite strong: X 1.6, so while the accelerated non-thermal electrons from this strong flare could produce a white light flare directly in this event, we still do not know whether most white light flares can be produced by the energy from the accelerated electrons, or whether this process works even in small flares. Therefore, further studies applying similar techniques to other flares that produce white light emission are needed to confirm whether they can also be produced directly or not.

\acknowledgments
Data are courtesy of the science teams of {\it Hinode}, {\it IRIS}, {\it RHESSI}, and {\it SDO}. {\it Hinode} is a Japanese mission developed and launched by ISAS/JAXA, with NAOJ as domestic partner and NASA and STFC (UK) as
international partners. It is operated by these agencies in cooperation with ESA and the NSC (Norway). {\it IRIS} is a NASA Small Explorer (SMEX) mission developed and operated by LMSAL with mission operations executed at NASA Ames Research center and major contributions to downlink communications funded by ESA and the Norwegian Space Centre. The {\it RHESSI} satellite is a NASA SMEX mission. HMI and AIA are instruments on board {\it SDO}, a mission for NASAÕs Living With a Star program. 
This work was supported by JSPS KAKENHI Grant Numbers JP25220703 (PI: S. Tsuneta), JP15K17622 and JP16H01187. The work of KSL was carried out by the joint research program of the Institute for Space-Earth Environmental Research (ISEE), Nagoya University. The work of SI was supported by JSPS KAKENHI Grant Number JP23340045, JP26287143, JPG2602, JP15H05816. The work of DHB was performed under contract with the Naval Research Laboratory and was funded by the NASA {\it Hinode} program.

\bibliographystyle{apj}
\bibliography{apj}

\begin{thebibliography}{}

\bibitem[\protect\citeauthoryear{{Alexander} \& {MacKinnon}}{{Alexander} \&
  {MacKinnon}}{1993}]{alexander&mackinnon_1993}
{Alexander}, D.,  \& {MacKinnon}, A.~L. 1993, \solphys, 144, 155

\bibitem[\protect\citeauthoryear{{Antonucci} et~al.}{{Antonucci}
  et~al.}{1982}]{antonucci_etal1982}
{Antonucci}, E., et~al. 1982, \solphys, 78, 107

\bibitem[\protect\citeauthoryear{{Bamba} et~al.}{{Bamba}
  et~al.}{2016}]{bamba_etal_prep}
{Bamba}, Y., {Lee}, K.-S., {Imada}, S.,  \& {Kusano}, K. 2016, in preparation

\bibitem[\protect\citeauthoryear{{Brosius}}{{Brosius}}{2003}]{brosius2003}
{Brosius}, J.~W. 2003, \apj, 586, 1417

\bibitem[\protect\citeauthoryear{{Brosius}}{{Brosius}}{2009}]{brosius2009}
{Brosius}, J.~W. 2009, \apj, 701, 1209

\bibitem[\protect\citeauthoryear{{Brosius}}{{Brosius}}{2013a}]{brosius2013a}
{Brosius}, J.~W. 2013a, \apj, 762, 133

\bibitem[\protect\citeauthoryear{{Brosius}}{{Brosius}}{2013b}]{brosius2013b}
{Brosius}, J.~W. 2013b, \apj, 777, 135

\bibitem[\protect\citeauthoryear{{Brown}}{{Brown}}{1971}]{brown_1971}
{Brown}, J.~C. 1971, \solphys, 18, 489

\bibitem[\protect\citeauthoryear{{Carmichael}}{{Carmichael}}{1964}]{carmichael_1964}
{Carmichael}, H. 1964, NASA Special Publication, 50, 451

\bibitem[\protect\citeauthoryear{{Carrington}}{{Carrington}}{1859}]{carrington_1859}
{Carrington}, R.~C. 1859, \mnras, 20, 13

\bibitem[\protect\citeauthoryear{{Chae}, {Sch{\"u}hle}, \& {Lemaire}}{{Chae}
  et~al.}{1998}]{chae_etal1998b}
{Chae}, J., {Sch{\"u}hle}, U.,  \& {Lemaire}, P. 1998, \apj, 505, 957

\bibitem[\protect\citeauthoryear{{Culhane} et~al.}{{Culhane}
  et~al.}{2007}]{culhane_etal2007a}
{Culhane}, J.~L., et~al. 2007, \solphys, 243, 19

\bibitem[\protect\citeauthoryear{{De Pontieu} et~al.}{{De Pontieu}
  et~al.}{2014}]{depontieu_etal2014}
{De Pontieu}, B., et~al. 2014, \solphys, 289, 2733

\bibitem[\protect\citeauthoryear{{del Zanna} et~al.}{{del Zanna}
  et~al.}{2006}]{delzanna_etal2006}
{del Zanna}, G., {Berlicki}, A., {Schmieder}, B.,  \& {Mason}, H.~E. 2006,
  \solphys, 234, 95

\bibitem[\protect\citeauthoryear{{Dere} et~al.}{{Dere}
  et~al.}{1997}]{dere_etal1997}
{Dere}, K.~P., {Landi}, E., {Mason}, H.~E., {Monsignori Fossi}, B.~C.,  \&
  {Young}, P.~R. 1997, \aaps, 125, 149

\bibitem[\protect\citeauthoryear{{Dere} \& {Mason}}{{Dere} \&
  {Mason}}{1993}]{dere&mason_1993}
{Dere}, K.~P.,  \& {Mason}, H.~E. 1993, \solphys, 144, 217

\bibitem[\protect\citeauthoryear{{Fisher}, {Canfield}, \& {McClymont}}{{Fisher}
  et~al.}{1985a}]{fisher_etal1985c}
{Fisher}, G.~H., {Canfield}, R.~C.,  \& {McClymont}, A.~N. 1985a, \apj, 289,
  434

\bibitem[\protect\citeauthoryear{{Fisher}, {Canfield}, \& {McClymont}}{{Fisher}
  et~al.}{1985b}]{fisher_etal1985b}
{Fisher}, G.~H., {Canfield}, R.~C.,  \& {McClymont}, A.~N. 1985b, \apj, 289,
  425

\bibitem[\protect\citeauthoryear{{Fisher}, {Canfield}, \& {McClymont}}{{Fisher}
  et~al.}{1985c}]{fisher_etal1985a}
{Fisher}, G.~H., {Canfield}, R.~C.,  \& {McClymont}, A.~N. 1985c, \apj, 289,
  414

\bibitem[\protect\citeauthoryear{{Fletcher} et~al.}{{Fletcher}
  et~al.}{2011}]{fletcher_etal2011}
{Fletcher}, L., et~al. 2011, \ssr, 159, 19

\bibitem[\protect\citeauthoryear{{Fletcher} \& {Hudson}}{{Fletcher} \&
  {Hudson}}{2008}]{fletcher_hudson2008}
{Fletcher}, L.,  \& {Hudson}, H.~S. 2008, \apj, 675, 1645

\bibitem[\protect\citeauthoryear{{Graham} \& {Cauzzi}}{{Graham} \&
  {Cauzzi}}{2015}]{graham&cauzzi_2015}
{Graham}, D.~R.,  \& {Cauzzi}, G. 2015, \apjl, 807, L22

\bibitem[\protect\citeauthoryear{{Hara} et~al.}{{Hara}
  et~al.}{2011}]{hara_etal2011}
{Hara}, H., {Watanabe}, T., {Harra}, L.~K., {Culhane}, J.~L.,  \& {Young},
  P.~R. 2011, \apj, 741, 107

\bibitem[\protect\citeauthoryear{{Hirayama}}{{Hirayama}}{1974}]{hirayama_1974}
{Hirayama}, T. 1974, \solphys, 34, 323

\bibitem[\protect\citeauthoryear{{Hudson}}{{Hudson}}{1972}]{hudson_1972}
{Hudson}, H.~S. 1972, \solphys, 24, 414

\bibitem[\protect\citeauthoryear{{Hudson} et~al.}{{Hudson}
  et~al.}{1992}]{hudson_etal1992}
{Hudson}, H.~S., {Acton}, L.~W., {Hirayama}, T.,  \& {Uchida}, Y. 1992, \pasj,
  44, L77

\bibitem[\protect\citeauthoryear{{Hudson}, {Canfield}, \& {Kane}}{{Hudson}
  et~al.}{1978}]{hudson_etal1978}
{Hudson}, H.~S., {Canfield}, R.~C.,  \& {Kane}, S.~R. 1978, \solphys, 60, 137

\bibitem[\protect\citeauthoryear{{Hudson}, {Wolfson}, \& {Metcalf}}{{Hudson}
  et~al.}{2006}]{hudson_etal2006}
{Hudson}, H.~S., {Wolfson}, C.~J.,  \& {Metcalf}, T.~R. 2006, \solphys, 234, 79

\bibitem[\protect\citeauthoryear{{Ichimoto} \& {Kurokawa}}{{Ichimoto} \&
  {Kurokawa}}{1984}]{ichimoto&kurokawa_1984}
{Ichimoto}, K.,  \& {Kurokawa}, H. 1984, \solphys, 93, 105

\bibitem[\protect\citeauthoryear{{Imada} et~al.}{{Imada}
  et~al.}{2008}]{imada_etal2008}
{Imada}, S., {Hara}, H., {Watanabe}, T., {Asai}, A., {Minoshima}, T., {Harra},
  L.~K.,  \& {Mariska}, J.~T. 2008, \apjl, 679, L155

\bibitem[\protect\citeauthoryear{{Imada}, {Murakami}, \& {Watanabe}}{{Imada}
  et~al.}{2015}]{imada_etal2015}
{Imada}, S., {Murakami}, I.,  \& {Watanabe}, T. 2015, Physics of Plasmas, 22,
  101206

\bibitem[\protect\citeauthoryear{{Imada} et~al.}{{Imada}
  et~al.}{2011}]{imada_etal2011}
{Imada}, S., {Murakami}, I., {Watanabe}, T., {Hara}, H.,  \& {Shimizu}, T.
  2011, \apj, 742, 70

\bibitem[\protect\citeauthoryear{{Kafatos} \& {Tucker}}{{Kafatos} \&
  {Tucker}}{1972}]{kafatos&tucker1972}
{Kafatos}, M.~C.,  \& {Tucker}, W.~H. 1972, \apj, 175, 837

\bibitem[\protect\citeauthoryear{{Kamio} et~al.}{{Kamio}
  et~al.}{2005}]{kamio_etal2005}
{Kamio}, S., {Kurokawa}, H., {Brooks}, D.~H., {Kitai}, R.,  \& {UeNo}, S. 2005,
  \apj, 625, 1027

\bibitem[\protect\citeauthoryear{{Kawate} \& {Imada}}{{Kawate} \&
  {Imada}}{2013}]{kawate&imada_2013}
{Kawate}, T.,  \& {Imada}, S. 2013, \apj, 775, 122

\bibitem[\protect\citeauthoryear{{Kleint} et~al.}{{Kleint}
  et~al.}{2016}]{kleint_etal2016}
{Kleint}, L., {Heinzel}, P., {Judge}, P.,  \& {Krucker}, S. 2016, \apj, 816, 88

\bibitem[\protect\citeauthoryear{{Kopp} \& {Pneuman}}{{Kopp} \&
  {Pneuman}}{1976}]{kopp&pneuman_1976}
{Kopp}, R.~A.,  \& {Pneuman}, G.~W. 1976, \solphys, 50, 85

\bibitem[\protect\citeauthoryear{{Krucker} et~al.}{{Krucker}
  et~al.}{2015}]{krucker_etal2015}
{Krucker}, S., et~al. 2015, \apj, 802, 19

\bibitem[\protect\citeauthoryear{{Kuhar} et~al.}{{Kuhar}
  et~al.}{2016}]{kuhar_etal2016}
{Kuhar}, M., {Krucker}, S., {Mart{\'{\i}}nez Oliveros}, J.~C., {Battaglia}, M.,
  {Kleint}, L., {Casadei}, D.,  \& {Hudson}, H.~S. 2016, \apj, 816, 6

\bibitem[\protect\citeauthoryear{{Landi} et~al.}{{Landi}
  et~al.}{2012}]{landi_etal2012}
{Landi}, E., {Del Zanna}, G., {Young}, P.~R., {Dere}, K.~P.,  \& {Mason}, H.~E.
  2012, \apj, 744, 99

\bibitem[\protect\citeauthoryear{{Leenaarts} et~al.}{{Leenaarts}
  et~al.}{2013}]{leenaarts_etal2013b}
{Leenaarts}, J., {Pereira}, T.~M.~D., {Carlsson}, M., {Uitenbroek}, H.,  \& {De
  Pontieu}, B. 2013, \apj, 772, 90

\bibitem[\protect\citeauthoryear{{Lemen} et~al.}{{Lemen}
  et~al.}{2012}]{lemen_etal2012}
{Lemen}, J.~R., et~al. 2012, \solphys, 275, 17

\bibitem[\protect\citeauthoryear{{Li}, {Ning}, \& {Zhang}}{{Li}
  et~al.}{2015}]{li_d_etal2015}
{Li}, D., {Ning}, Z.,  \& {Zhang}, Q. 2015, ArXiv e-prints

\bibitem[\protect\citeauthoryear{{Li} \& {Ding}}{{Li} \&
  {Ding}}{2011}]{li&ding_2011}
{Li}, Y.,  \& {Ding}, M.~D. 2011, \apj, 727, 98

\bibitem[\protect\citeauthoryear{{Li} et~al.}{{Li}
  et~al.}{2015}]{li_y_etal2015}
{Li}, Y., {Ding}, M.~D., {Qiu}, J.,  \& {Cheng}, J.~X. 2015, \apj, 811, 7

\bibitem[\protect\citeauthoryear{{Lin} et~al.}{{Lin}
  et~al.}{2002}]{lin_etal2002}
{Lin}, R.~P., et~al. 2002, \solphys, 210, 3

\bibitem[\protect\citeauthoryear{{Machado}, {Emslie}, \& {Avrett}}{{Machado}
  et~al.}{1989}]{machado_etal1989}
{Machado}, M.~E., {Emslie}, A.~G.,  \& {Avrett}, E.~H. 1989, \solphys, 124, 303

\bibitem[\protect\citeauthoryear{{Mariska}}{{Mariska}}{1992}]{mariska_1992_b}
{Mariska}, J.~T. 1992, {The solar transition region} (Cambridge University
  Press)

\bibitem[\protect\citeauthoryear{{Mart{\'{\i}}nez-Sykora}
  et~al.}{{Mart{\'{\i}}nez-Sykora} et~al.}{2016}]{martinez_etal2016}
{Mart{\'{\i}}nez-Sykora}, J., {De Pontieu}, B., {Hansteen}, V.~H.,  \&
  {Gudiksen}, B. 2016, \apj, 817, 46

\bibitem[\protect\citeauthoryear{{Matthews} et~al.}{{Matthews}
  et~al.}{2003}]{matthews_etal2003}
{Matthews}, S.~A., {van Driel-Gesztelyi}, L., {Hudson}, H.~S.,  \& {Nitta},
  N.~V. 2003, \aap, 409, 1107

\bibitem[\protect\citeauthoryear{{Metcalf} et~al.}{{Metcalf}
  et~al.}{2003}]{metcalf_etal2003}
{Metcalf}, T.~R., {Alexander}, D., {Hudson}, H.~S.,  \& {Longcope}, D.~W. 2003,
  \apj, 595, 483

\bibitem[\protect\citeauthoryear{{Milligan} \& {Dennis}}{{Milligan} \&
  {Dennis}}{2009}]{milligan&dennis_2009}
{Milligan}, R.~O.,  \& {Dennis}, B.~R. 2009, \apj, 699, 968

\bibitem[\protect\citeauthoryear{{Milligan} et~al.}{{Milligan}
  et~al.}{2014}]{milligan_etal2014}
{Milligan}, R.~O., et~al. 2014, \apj, 793, 70

\bibitem[\protect\citeauthoryear{{Nagai}}{{Nagai}}{1980}]{nagai_1980}
{Nagai}, F. 1980, \solphys, 68, 351

\bibitem[\protect\citeauthoryear{{Nagai} \& {Emslie}}{{Nagai} \&
  {Emslie}}{1984}]{nagai&emslie_1984}
{Nagai}, F.,  \& {Emslie}, A.~G. 1984, \apj, 279, 896

\bibitem[\protect\citeauthoryear{{Neidig}}{{Neidig}}{1989}]{neidig_1989}
{Neidig}, D.~F. 1989, \solphys, 121, 261

\bibitem[\protect\citeauthoryear{{Neupert}}{{Neupert}}{1968}]{neupert_1968}
{Neupert}, W.~M. 1968, \apjl, 153, L59

\bibitem[\protect\citeauthoryear{{Nitta}, {Imada}, \& {Yamamoto}}{{Nitta}
  et~al.}{2012}]{nitta_etal2012}
{Nitta}, S., {Imada}, S.,  \& {Yamamoto}, T.~T. 2012, \solphys, 276, 183

\bibitem[\protect\citeauthoryear{{Olluri}, {Gudiksen}, \& {Hansteen}}{{Olluri}
  et~al.}{2013}]{olluri_etal2013}
{Olluri}, K., {Gudiksen}, B.~V.,  \& {Hansteen}, V.~H. 2013, \apj, 767, 43

\bibitem[\protect\citeauthoryear{{Pereira} et~al.}{{Pereira}
  et~al.}{2015}]{pereira_2015}
{Pereira}, T.~M.~D., {Carlsson}, M., {De Pontieu}, B.,  \& {Hansteen}, V. 2015,
  \apj, 806, 14

\bibitem[\protect\citeauthoryear{{Phillips}, {Feldman}, \& {Landi}}{{Phillips}
  et~al.}{2008}]{phillips_etal2008_b}
{Phillips}, K.~J.~H., {Feldman}, U.,  \& {Landi}, E. 2008, {Ultraviolet and
  X-ray Spectroscopy of the Solar Atmosphere} (Cambridge University Press)

\bibitem[\protect\citeauthoryear{{Polito} et~al.}{{Polito}
  et~al.}{2016}]{polito_etal2016}
{Polito}, V., {Reep}, J.~W., {Reeves}, K.~K., {Sim{\~o}es}, P.~J.~A.,
  {Dud{\'{\i}}k}, J., {Del Zanna}, G., {Mason}, H.~E.,  \& {Golub}, L. 2016,
  \apj, 816, 89

\bibitem[\protect\citeauthoryear{{Polito} et~al.}{{Polito}
  et~al.}{2015}]{polito_etal2015}
{Polito}, V., {Reeves}, K.~K., {Del Zanna}, G., {Golub}, L.,  \& {Mason}, H.~E.
  2015, \apj, 803, 84

\bibitem[\protect\citeauthoryear{{Reep} \& {Russell}}{{Reep} \&
  {Russell}}{2016}]{reep&russell_2016}
{Reep}, J.~W.,  \& {Russell}, A.~J.~B. 2016, \apjl, 818, L20

\bibitem[\protect\citeauthoryear{{Schou} et~al.}{{Schou}
  et~al.}{2012}]{schou_etal2012}
{Schou}, J., et~al. 2012, \solphys, 275, 229

\bibitem[\protect\citeauthoryear{{Sturrock}}{{Sturrock}}{1968}]{sturrock_1968}
{Sturrock}, P.~A. 1968, The Astronomical Journal Supplement, 73, 78

\bibitem[\protect\citeauthoryear{{Thalmann} et~al.}{{Thalmann}
  et~al.}{2015}]{thalmann_etal2015}
{Thalmann}, J.~K., {Su}, Y., {Temmer}, M.,  \& {Veronig}, A.~M. 2015, \apjl,
  801, L23

\bibitem[\protect\citeauthoryear{{Tian} et~al.}{{Tian}
  et~al.}{2015}]{tian_etal2015}
{Tian}, H., {Young}, P.~R., {Reeves}, K.~K., {Chen}, B., {Liu}, W.,  \&
  {McKillop}, S. 2015, \apj, 811, 139

\bibitem[\protect\citeauthoryear{{{\v S}vestka}}{{{\v
  S}vestka}}{1966}]{svestka_1966}
{{\v S}vestka}, Z. 1966, \ssr, 5, 388

\bibitem[\protect\citeauthoryear{{Veronig} \& {Polanec}}{{Veronig} \&
  {Polanec}}{2015}]{veronig&polanec_2015}
{Veronig}, A.~M.,  \& {Polanec}, W. 2015, \solphys, 290, 2923

\bibitem[\protect\citeauthoryear{{Watanabe} et~al.}{{Watanabe}
  et~al.}{2010a}]{watanabe_etal2010}
{Watanabe}, K., {Krucker}, S., {Hudson}, H., {Shimizu}, T., {Masuda}, S.,  \&
  {Ichimoto}, K. 2010a, \apj, 715, 651

\bibitem[\protect\citeauthoryear{{Watanabe} et~al.}{{Watanabe}
  et~al.}{2010b}]{watanabe_t_etal2010}
{Watanabe}, T., {Hara}, H., {Sterling}, A.~C.,  \& {Harra}, L.~K. 2010b, \apj,
  719, 213

\bibitem[\protect\citeauthoryear{{Yokoyama} \& {Shibata}}{{Yokoyama} \&
  {Shibata}}{2001}]{yokoyama&shibata_2001}
{Yokoyama}, T.,  \& {Shibata}, K. 2001, \apj, 549, 1160

\bibitem[\protect\citeauthoryear{{Young}}{{Young}}{2015}]{young_2015}
{Young}, P.~R. 2015, ArXiv e-prints

\bibitem[\protect\citeauthoryear{{Young} et~al.}{{Young}
  et~al.}{2007}]{young_etal2007a}
{Young}, P.~R., et~al. 2007, \pasj, 59, 857

\bibitem[\protect\citeauthoryear{{Zirin} \& {Neidig}}{{Zirin} \&
  {Neidig}}{1981}]{zirin&neidig_1981}
{Zirin}, H.,  \& {Neidig}, D.~F. 1981, \apjl, 248, L45

\end{thebibliography}


\clearpage
\begin{deluxetable}{c c c}
\tablecaption{List of the Spectral Lines Used in the Present Study \label{tbl-1}} \tablewidth{0pt}
\tablehead{\colhead{Instrument} & \colhead{Line ID ($\rm{\AA}$)} & \colhead{$\log T_{max}$ (K)}}
	\startdata
 	EIS 	& \ion{He}{2} 256.32 			& 4.9 	\nl
		& \ion{O}{5} 248.46			& 5.4		\nl
		& \ion{Fe}{10} 184.54		& 6.1		\nl
		& \ion{Fe}{12} 195.12		& 6.2		\nl
		& \ion{Fe}{14} 264.79 $\ast$	& 6.3		\nl
		& \ion{Fe}{14} 274.20 $\ast$	& 6.3		\nl
		& \ion{Fe}{15} 284.16 		& 6.4		\nl
		& \ion{Fe}{16} 263.00		& 6.8 	\nl
		& \ion{Fe}{23} 263.77		& 7.2		\nl
		& \ion{Fe}{24} 192.03		& 7.2		\nl
	IRIS	& \ion{O}{1} 1355.6			& 4.5		\nl
		& \ion{Si}{4} 1402.8			& 4.9		\nl
		& \ion{Fe}{21} 1354.1		& 7.1		\nl
		& \ion{Mg}{2} 2798.8			& 4.0		\nl
		& \ion{O}{4} 1399.77	$\ast$	& 5.2		\nl
		& \ion{O}{4} 1401.16	$\ast$	& 5.2 	\nl
	\enddata
        \tablecomments{$ $ The peak formation temperatures of the spectral lines are taken from the Chianti database version 7.0. Lines used for the density determination are marked with asterisks. }	
\end{deluxetable}


\clearpage
\begin{figure*}
\epsscale{1.8}
\plotone{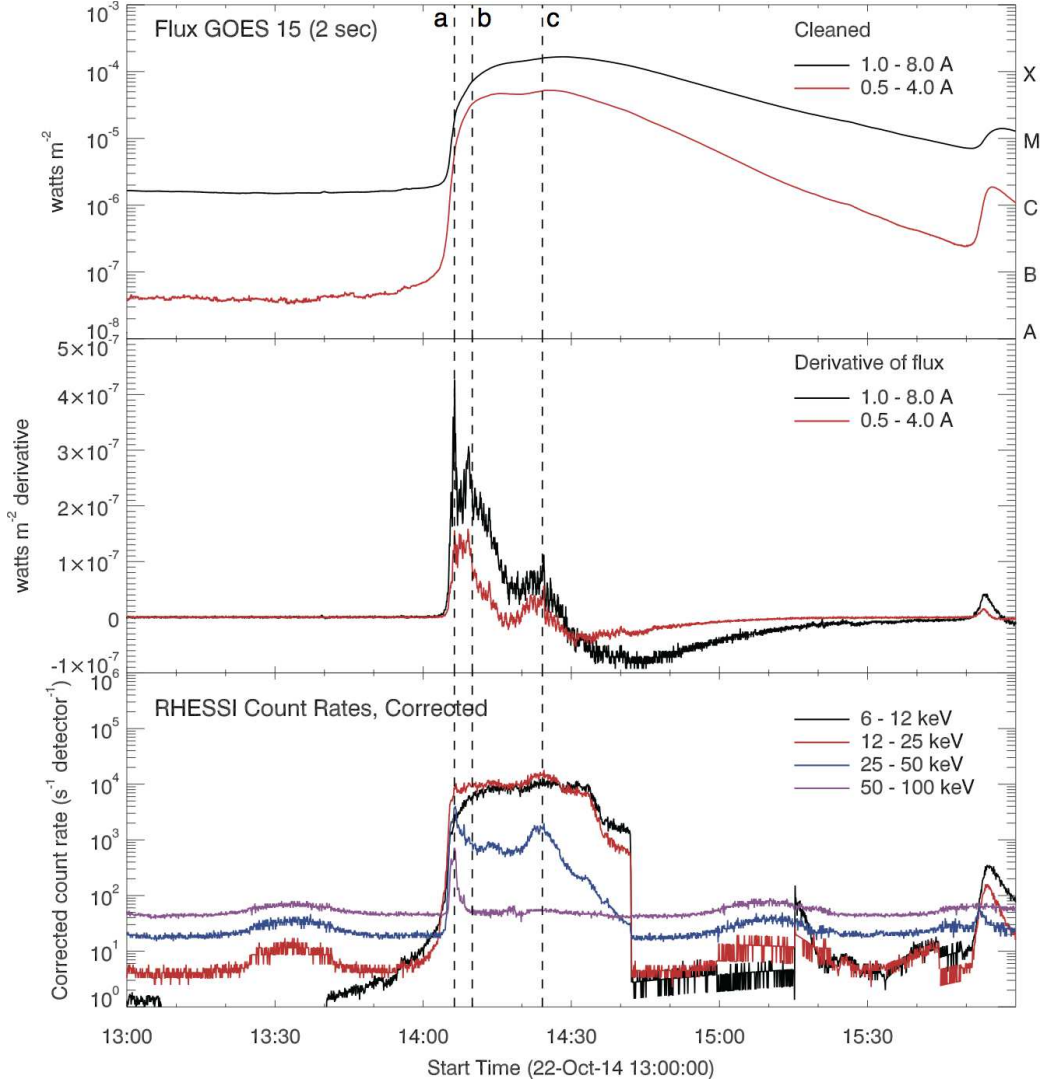} \caption{{\it GOES} x-ray light curve (top) of the X1.6 flare on 2014 October 22 14:06 UT and its time derivative (middle). The {\it RHESSI} count rates for the different energy bands are plotted in the bottom panel. The vertical dashed lines (a-c) mark the times of the three peaks in the time derivative of the {\it GOES} X-ray light curve.  \label{fig1}}
\end{figure*}


\clearpage
\begin{figure*}
\epsscale{1.8}
\plotone{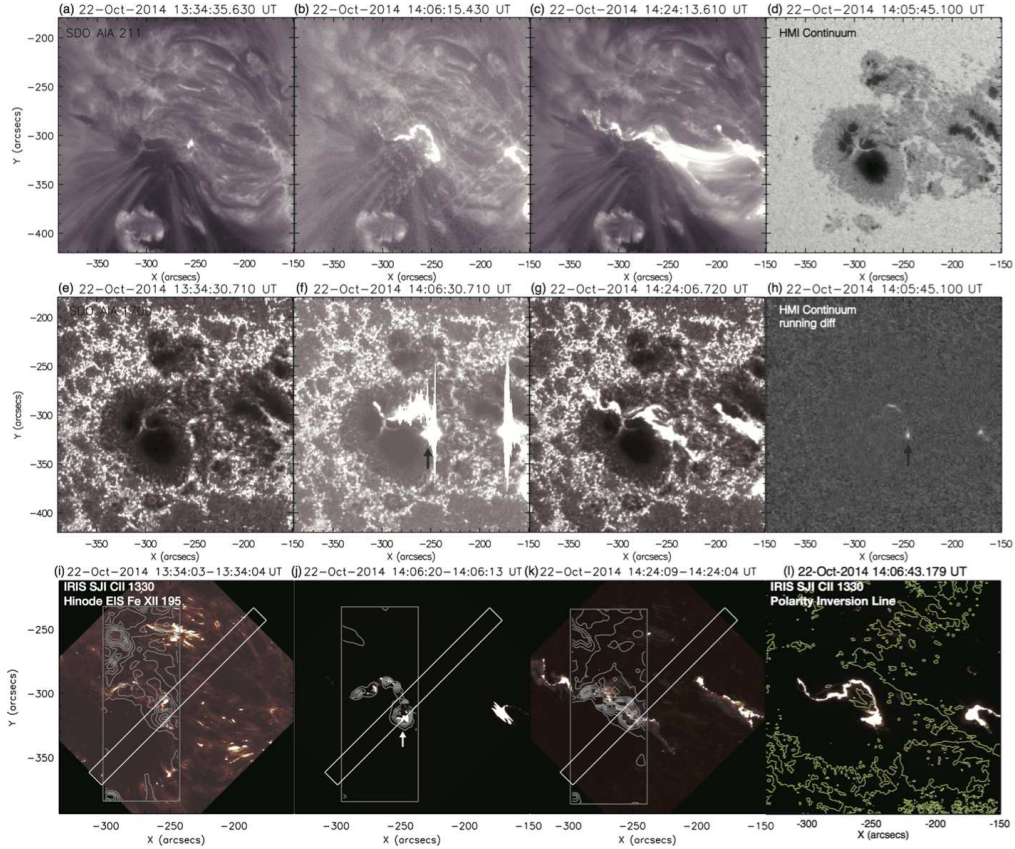} \caption{Context {\it SDO} and {\it IRIS}/SJI \ion{C}{2} 1330$~\rm{\AA}$ images for the X1.6 class flare on 2014 October 22. (a-c): AIA 211$~\rm{\AA}$, (e-g): AIA 1700$~\rm{\AA}$ channel images,  (i-k): {\it IRIS} \ion{C}{2} 1330$~\rm{\AA}$ slit jaw images overlaid with EIS 195$~\rm{\AA}$ intensity contours (gray line) taken before the flare ($\sim$13:37 UT), at the time of the first HXR peak ($\sim$14:06 UT), and at the flare peak ($\sim$14:24 UT). The white dashed box in the {\it IRIS} SJI corresponds to the location of the {\it IRIS} raster. Panels (d) and (h) display the {\it SDO}/HMI continuum and a running difference image at 14:05:45 UT, respectively. Panel (l) depicts the polarity inversion line from the HMI magnetogram on the {\it IRIS} SJI around timing (a) in Figure 1. The bright kernel we analyzed is marked with an arrow.  \label{fig2}}
\end{figure*}



\clearpage
\begin{figure*}
\epsscale{1.8}
\plotone{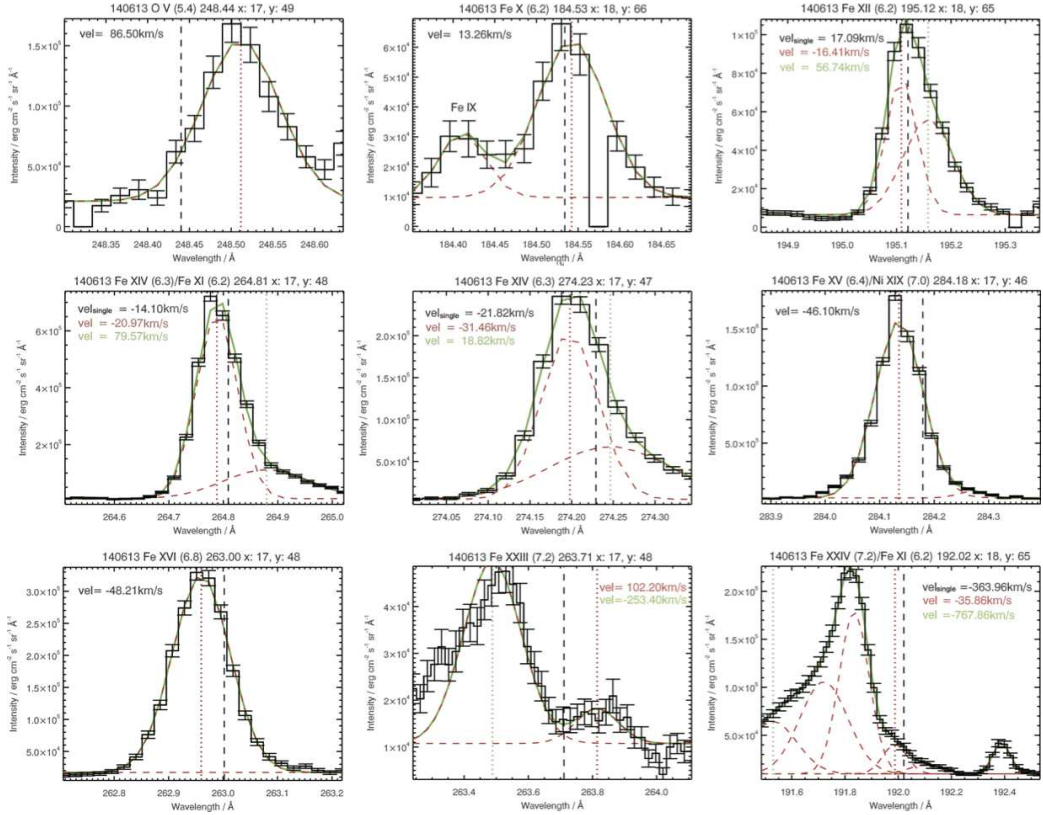}\caption{EIS spectral line profiles of each emission line for the bright kernel observed at 14:06:13 (impulsive phase). The green solid lines and red dashed lines represent the total line profiles from single or multiple Gaussian fits and each Gaussian component, respectively. The black vertical dashed lines represent the rest wavelengths obtained from averaging the line center before the flare from 13:01 to 13:51 UT.  The red green vertical dotted lines correspond to the line centers of the fitted line profile for the single gaussian and the shifted component from the multiple gaussian fit.  \label{fig3}}
\end{figure*}

\clearpage
\begin{figure*}
\epsscale{1.8}
\plotone{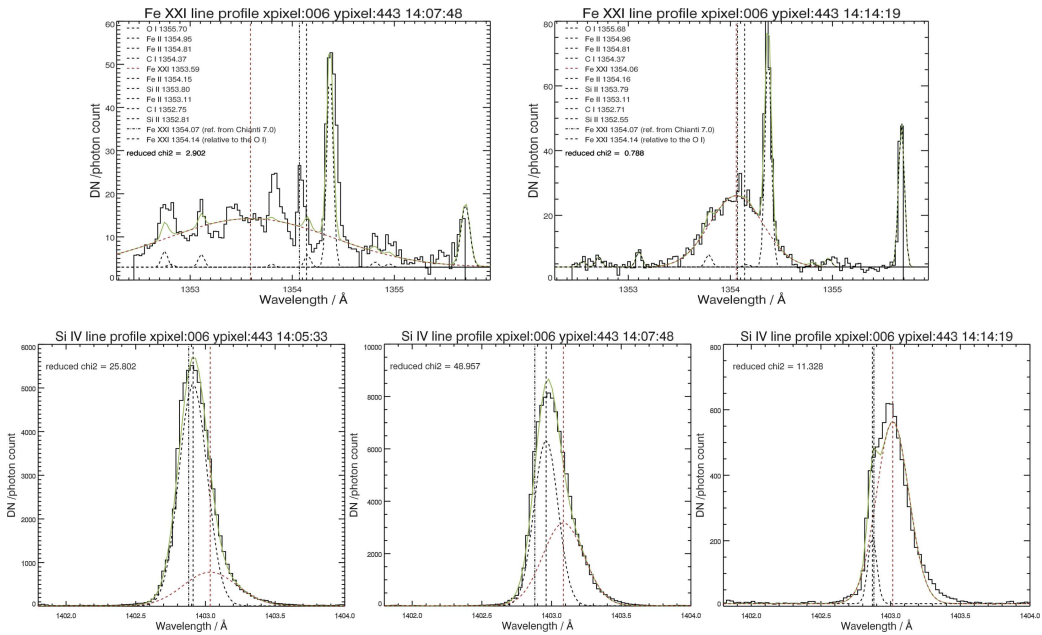} \caption{{\it IRIS} spectral line profiles of \ion{Fe}{21} (top) and \ion{Si}{4} (bottom) for the bright kernel observed during the impulsive phase. The green solid lines and black dotted lines represent the total line profiles from multiple Gaussian fits and each Gaussian component, respectively. In the upper panels, the black vertical dashed and dot-dashed lines represent the rest wavelengths relative to \ion{O}{1} and the reference wavelengths from the {\it Chianti} atomic database, respectively. In the lower panels, the dot-dashed and dashed lines show the rest wavelength and the first gaussian fit component of \ion{Si}{4}, respectively. The red vertical dashed line corresponds to the line center of the fitted line profile for the shifted component of \ion{Fe}{21} and \ion{Si}{4}. The \ion{Fe}{21} emission and the second Gaussian fit for \ion{Si}{4} are marked with a red dotted Gaussian line profile. \label{fig4} }
\end{figure*}


\clearpage

\begin{figure*}
\epsscale{1.8}
\plotone{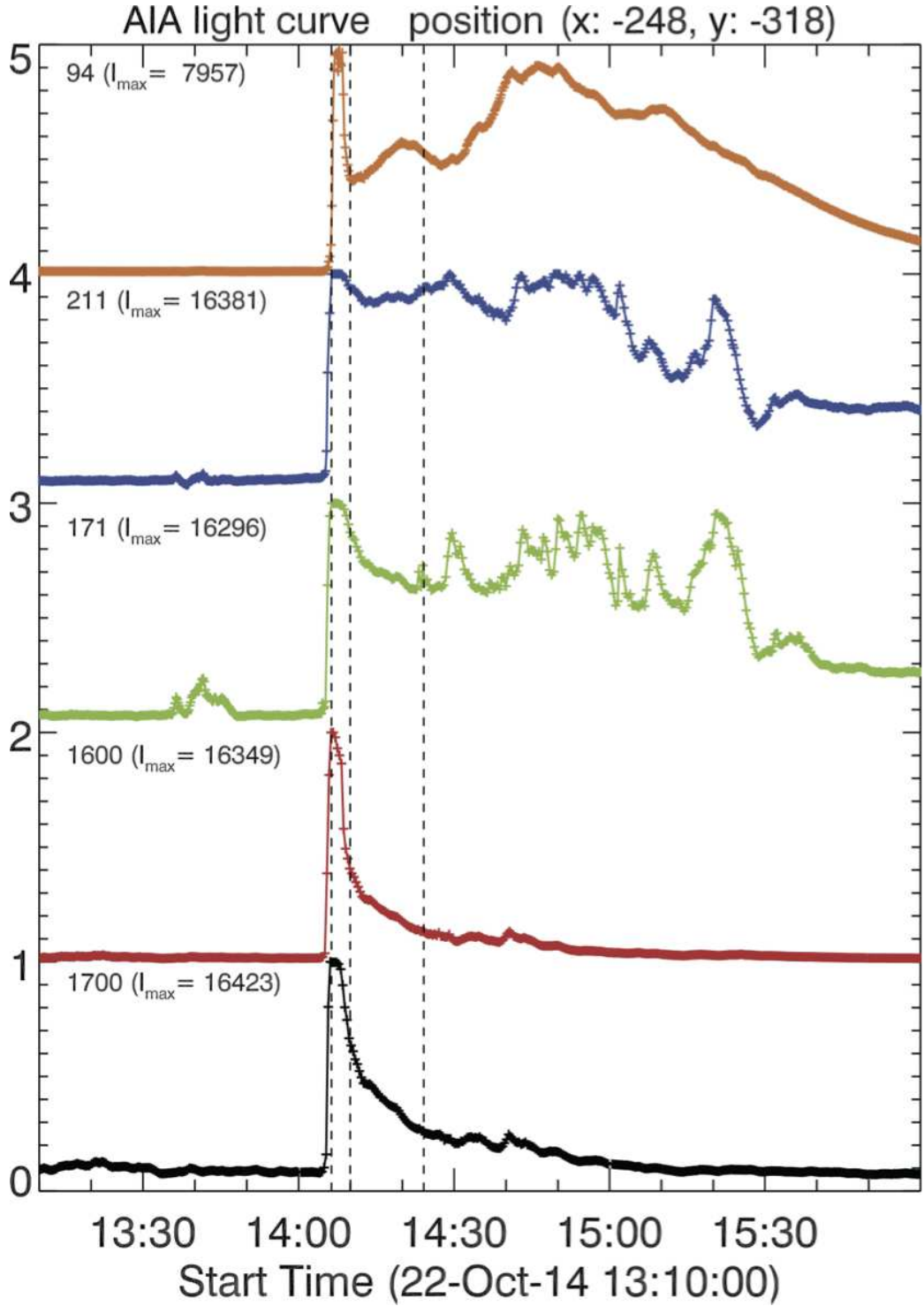} \caption{{\it SDO}/AIA light curves for the bright kernel. Intensities are normalized by the maximum intensity during the flare observation. The dashed vertical lines correspond to the same times marked in Figure 1. \label{fig5}}
\end{figure*}




\clearpage

\begin{figure*}
\epsscale{1.8}
\plotone{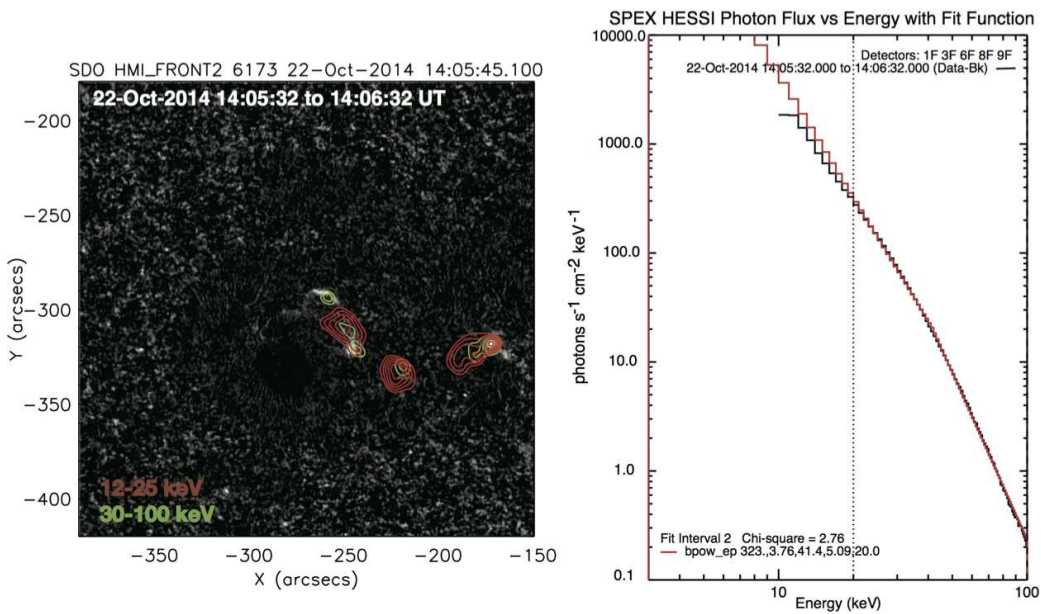} \caption{Left: HMI continuum difference image with HXR (30-100 keV) and SXR (12-25 keV) contours overlaid from the {\it RHESSI} cleaned image covering 14:05:32-14:06:32 UT. Red and green contours correspond to 50, 60, 70, 80, and 90$\%$ of the HXR and SXR intensity, respectively. Right: Fitted {\it RHESSI} spectrum with energy for a flux in the energy range of 20-100 keV during 14:05:32-14:06:32 UT. Black and red lines indicate the observed and fitted spectrum, respectively.  \label{fig6}}
\end{figure*}

\clearpage
\begin{figure*}
\epsscale{1.8}
\plotone{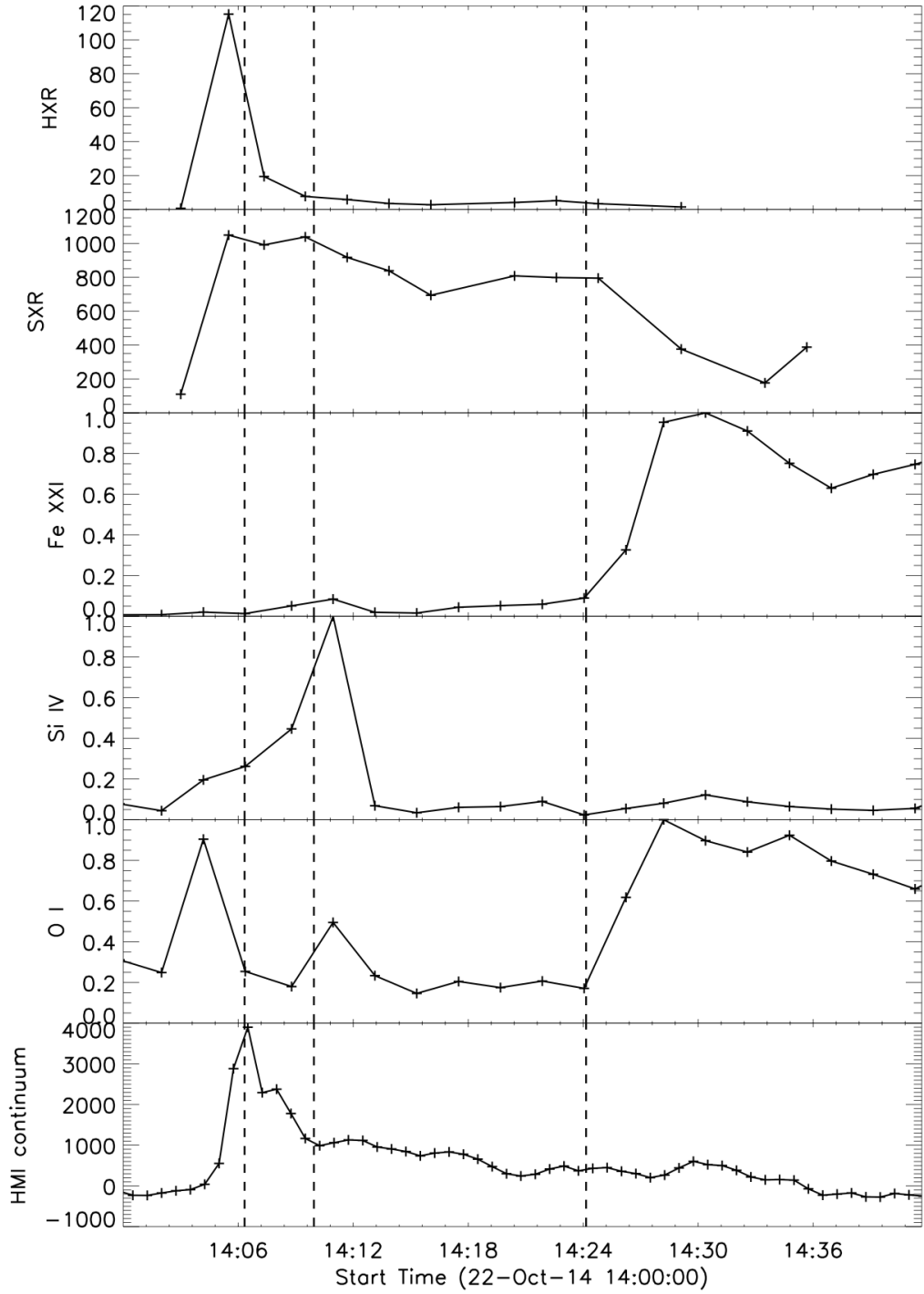} \caption{Light curves for the multi-wavelength observations: (a) {\it RHESSI} HXR, (b) {\it RHESSI} SXR, (c) IRIS \ion{Fe}{21}, (d) \ion{Si}{4}, (e) \ion{O}{1}, and (f) SDO/HMI continuum. The intensities from the IRIS spectra are normalized by the maximum intensity during the flare. The HMI continuum intensity peak is enhanced by around $3.8 \times 10^4$ erg $\rm{cm^{-2} ~ s^{-1} ~ sr^{-1}}$.    \label{fig7}}
\end{figure*}

\clearpage
\begin{figure*}
\epsscale{1.8}
\plotone{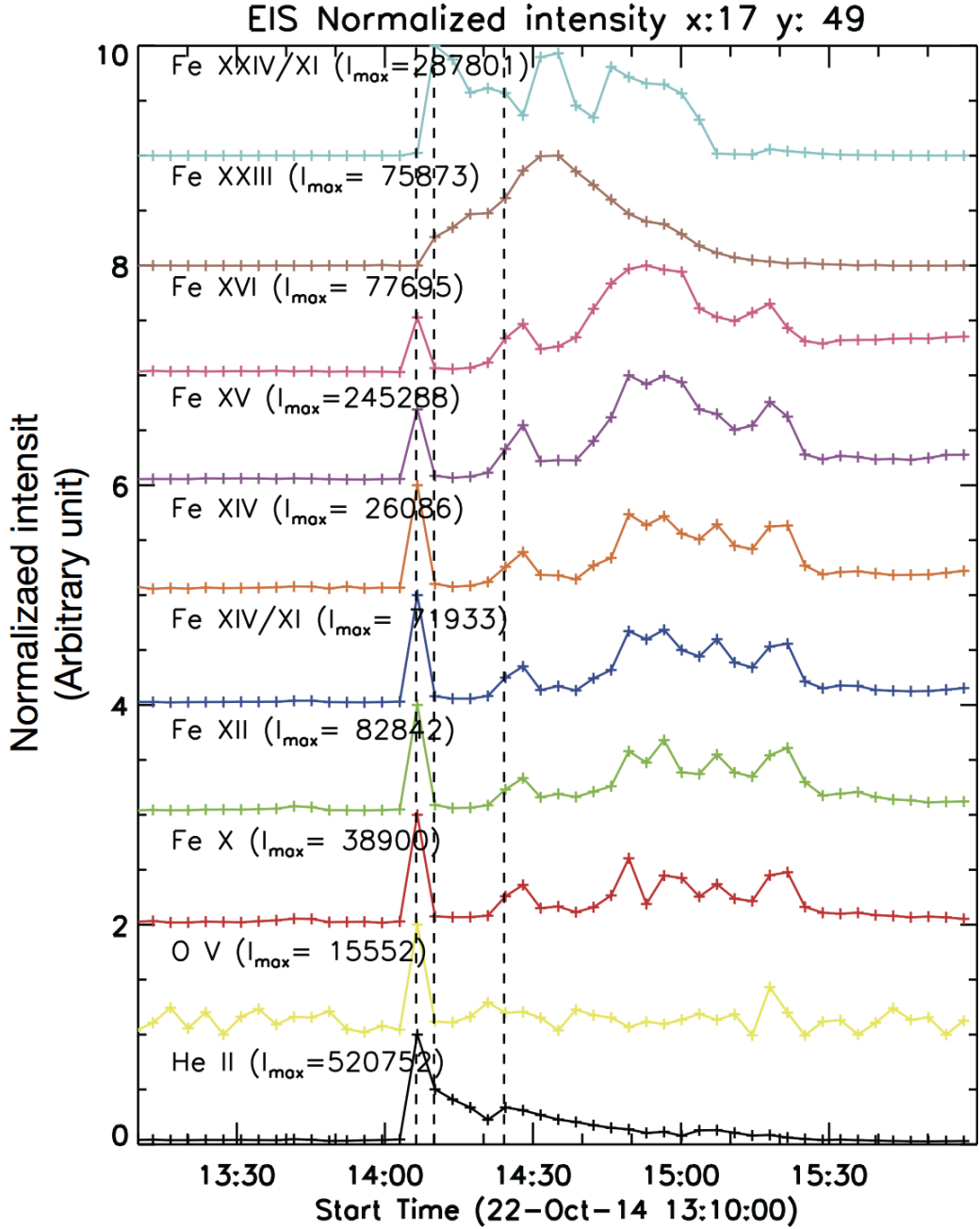} \caption{Temporal variation of the EIS spectral line intensities for the bright kernel. Intensities are normalized by the maximum intensity during the flare observation. The vertical dashed lines correspond to the times marked in Figure 1. \label{fig8}} 
\end{figure*}

\clearpage
\begin{figure*}
\epsscale{1.8}
\plotone{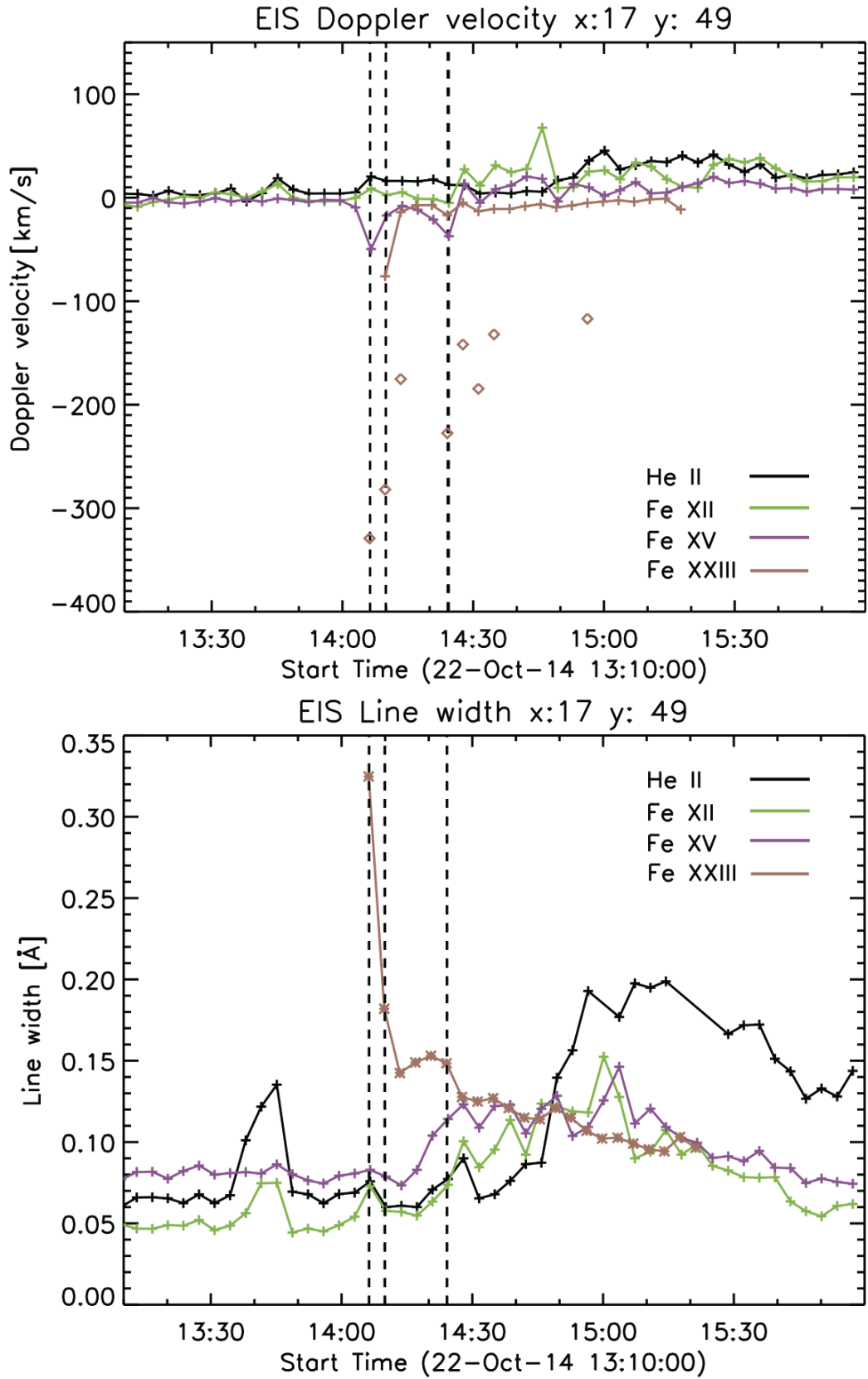}\caption{Temporal variation of the Doppler velocity (top) and line width (bottom) for the bright kernel from the EIS spectral lines, \ion{He}{2}, \ion{Fe}{12}, \ion{Fe}{15}, and \ion{Fe}{23}. The vertical dashed lines correspond to the same times as shown in Figure 1. \label{fig9}}
\end{figure*}

\clearpage
\begin{figure*}
\epsscale{1.8}
\plotone{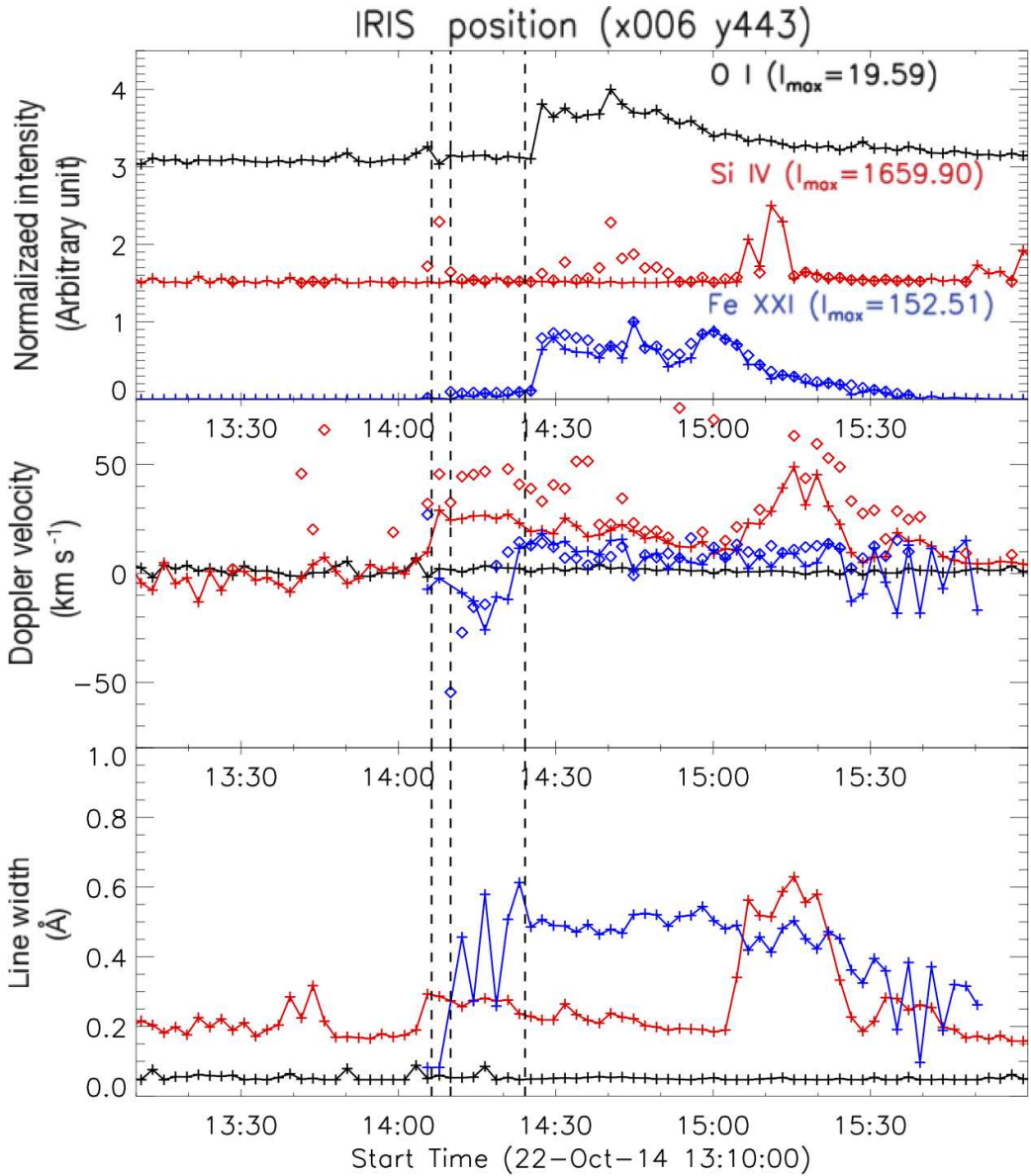} \caption{Temporal variation of the intensity (top), Doppler velocity (middle), and line width (bottom) for the bright kernel from the {\it IRIS} spectral lines, \ion{O}{1} (black), \ion{Si}{4} (red), and \ion{Fe}{21} (blue). Intensities are normalized by the maximum intensity during the flare observation. The vertical dashed lines correspond to the same times as shown in Figure 1. \label{fig10}}
\end{figure*}




\clearpage
\begin{figure*}
\epsscale{1.}
\plotone{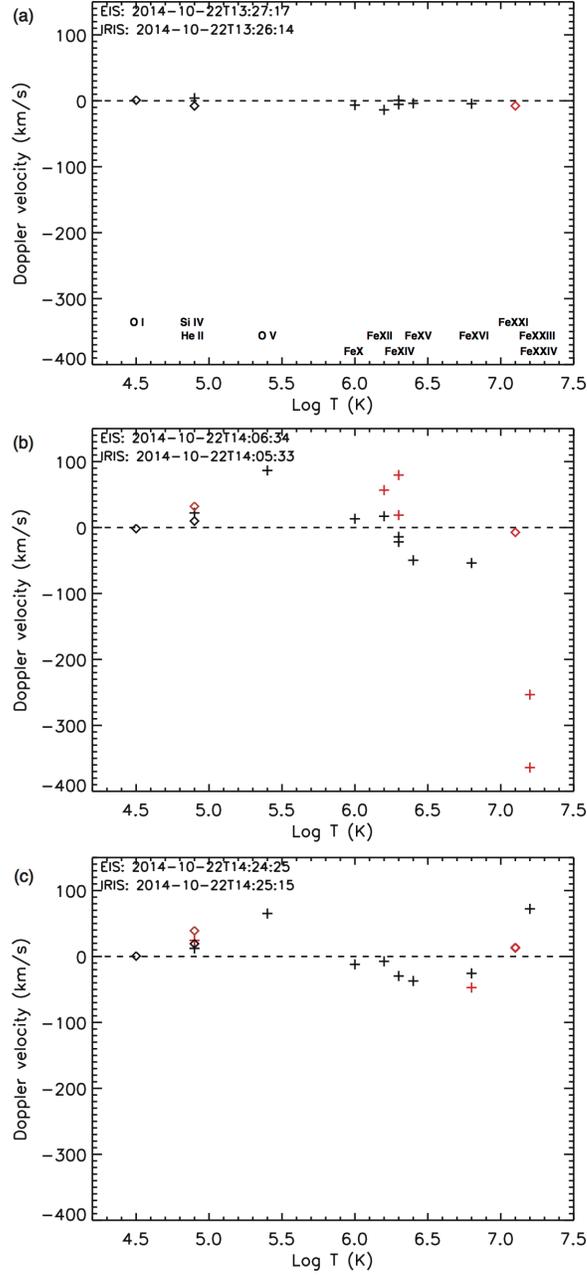}\caption{Doppler velocities from EIS and {\it IRIS} as a function of the peak formation temperature for the bright kernel at different times, (a) before the flare, (b) the impulsive phase, and (c) the gradual phase.  Diamonds indicate the Doppler velocities from the {\it IRIS} spectra while crosses represent the Doppler velocities from the EIS spectra. Black and red indicate the velocities calculated from the single and multiple Gaussian components relative to the rest wavelengths. The element and ionization information of the spectral lines are noted at the bottom of the panel (a). \label{fig11}}
\end{figure*}

\clearpage
\begin{figure*}
\epsscale{1.8}
\plotone{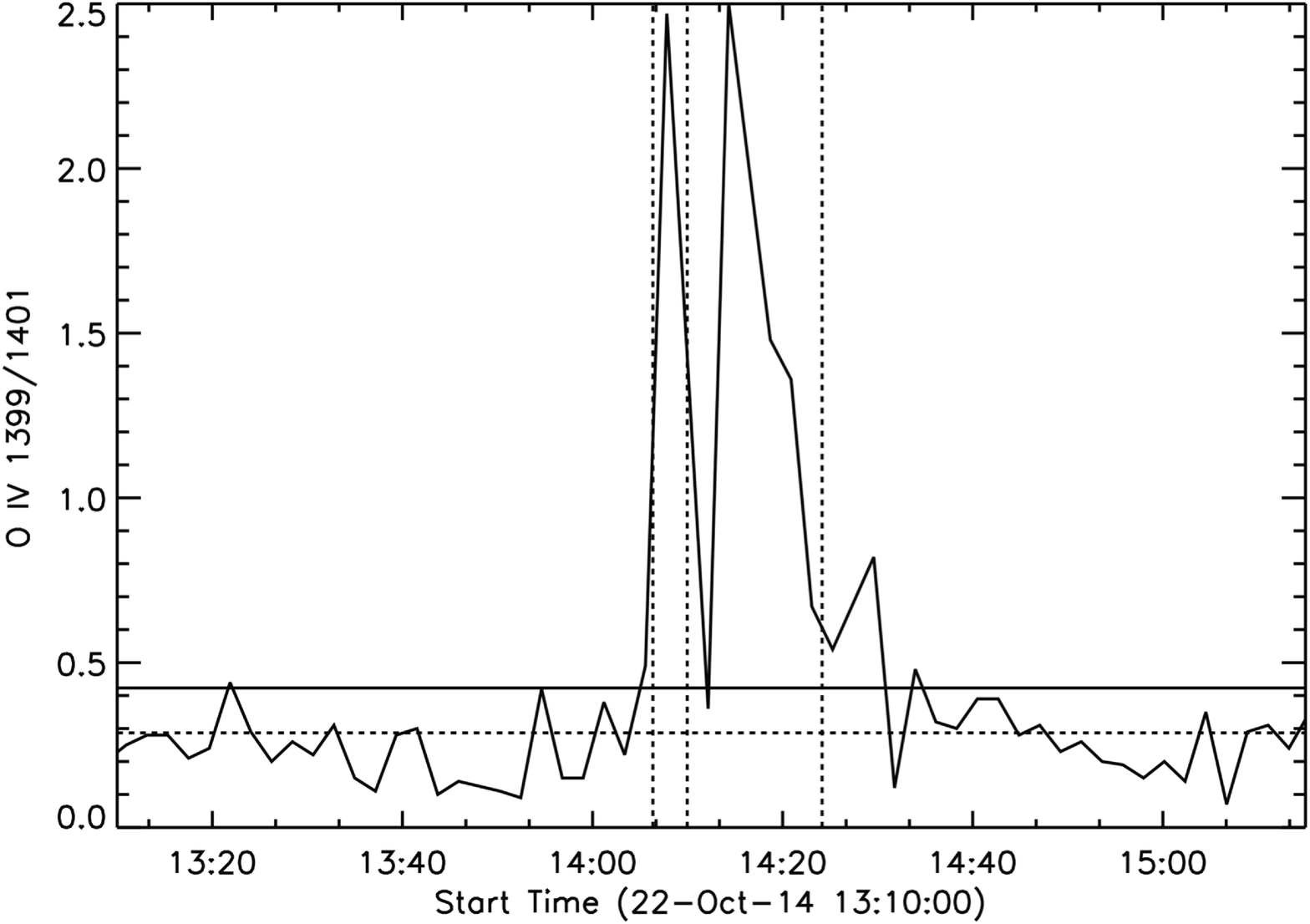}
\plotone{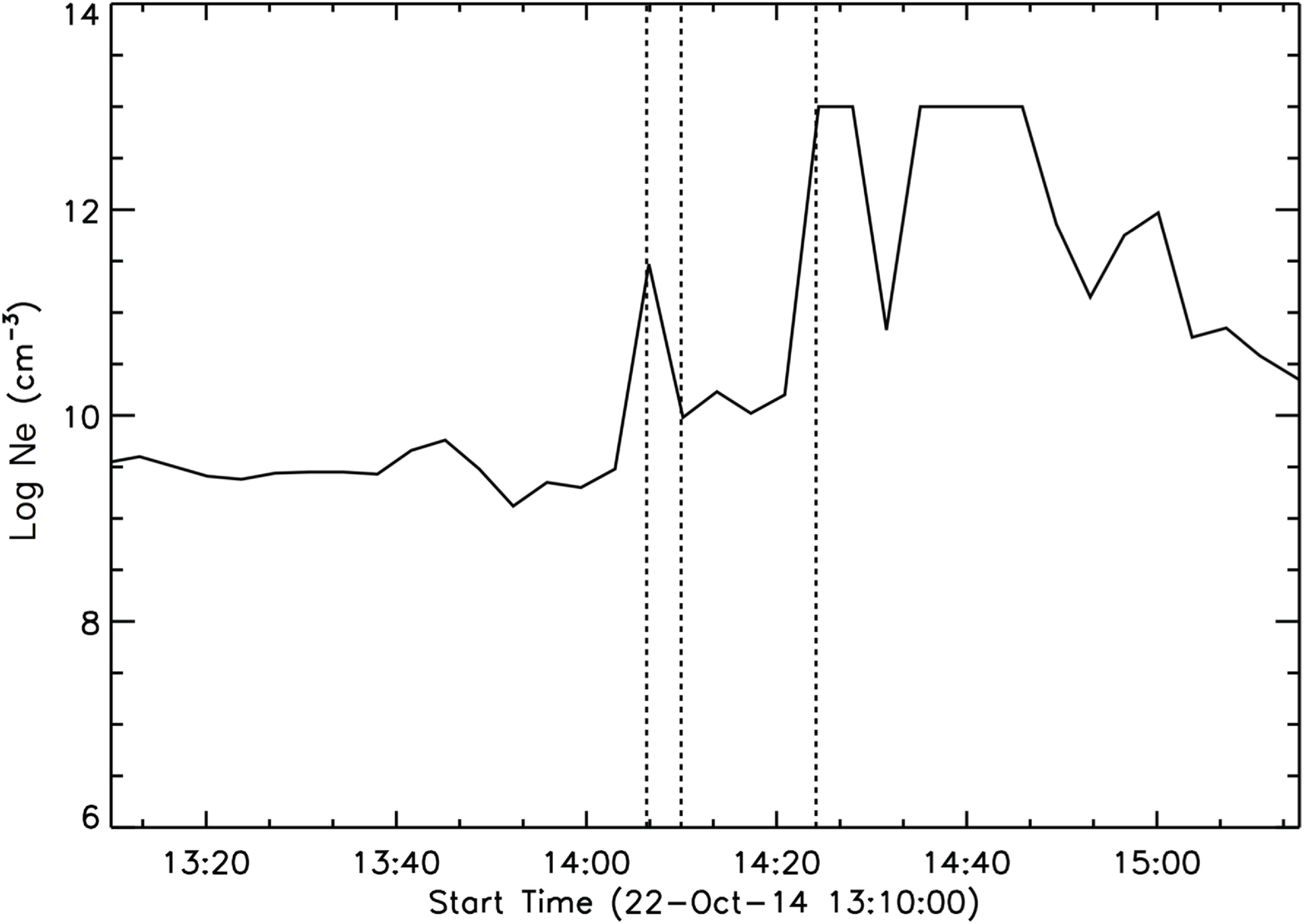}
\caption{Top: temporal variation of the {\it IRIS} \ion{O}{4} 1399.8 $\rm{\AA}$ / 1401.2 $\rm{\AA}$ intensity ratio for the bright kernel. The horizontal solid and dashed lines indicate the high density limit ratio of 0.43 (log $N_{e}$=13) and the averaged intensity ratio before and after the impulsive phase, respectively. Bottom: temporal variation of the density measured from the EIS \ion{Fe}{14} 264.79 $\rm{\AA}$ / 274.20 $\rm{\AA}$ intensity ratio for the bright kernel. The vertical dashed lines indicate the times marked in Figure 1. \label{fig12}}
\end{figure*}

\clearpage
\begin{figure*}
\epsscale{1.8}
\plotone{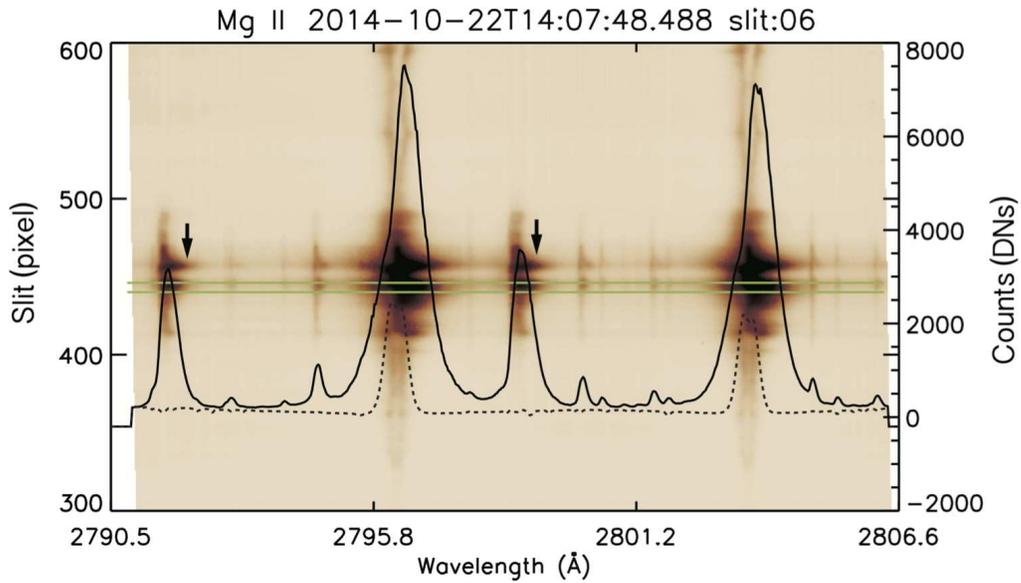} \caption{{\it IRIS} detector images of the \ion{Mg}{2} h \& k spectral windows overlaid with the averaged spectral line profiles at the location marked by the two horizontal green lines. The solid and dotted lines represent the line profiles during the impulsive phase and before the flare (around 12:00 UT), respectively. The arrows indicate the red shifts in the \ion{Mg}{2} spectral lines. \label{fig13}}
\end{figure*}

\clearpage
\begin{figure*}
\epsscale{1.8}
\plotone{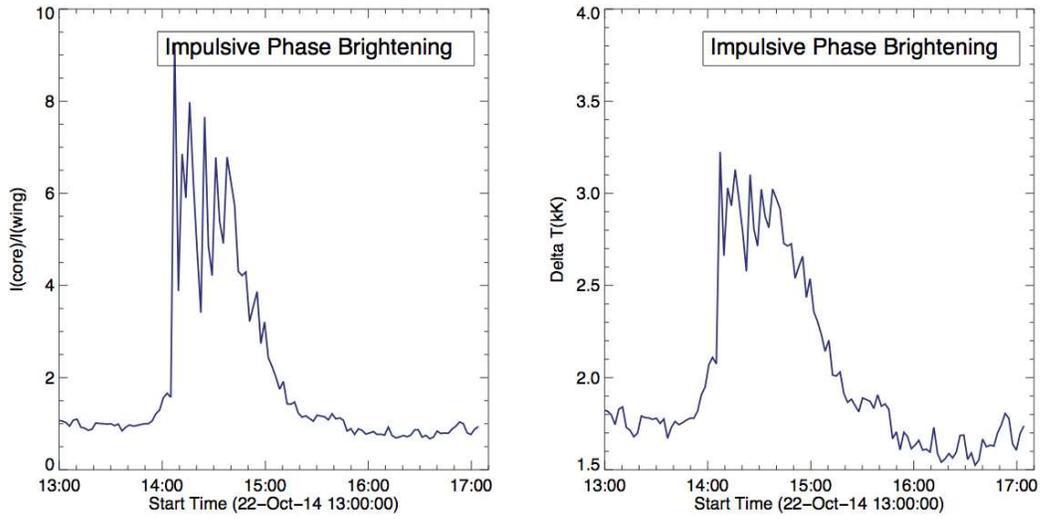} \caption{Left: Temporal variation of the line core to wing intensity ratio of the two blended \ion{Mg}{2} triplet lines, 2798.75 $\rm{\AA}$ and 2798.82 $\rm{\AA}$, for the bright kernel. Right: Temporal variation of the temperature changes ($\Delta T$) between the line core and wing formation regions for the bright kernel. \label{fig14}}
\end{figure*}

\clearpage
\begin{figure*}
\plotone{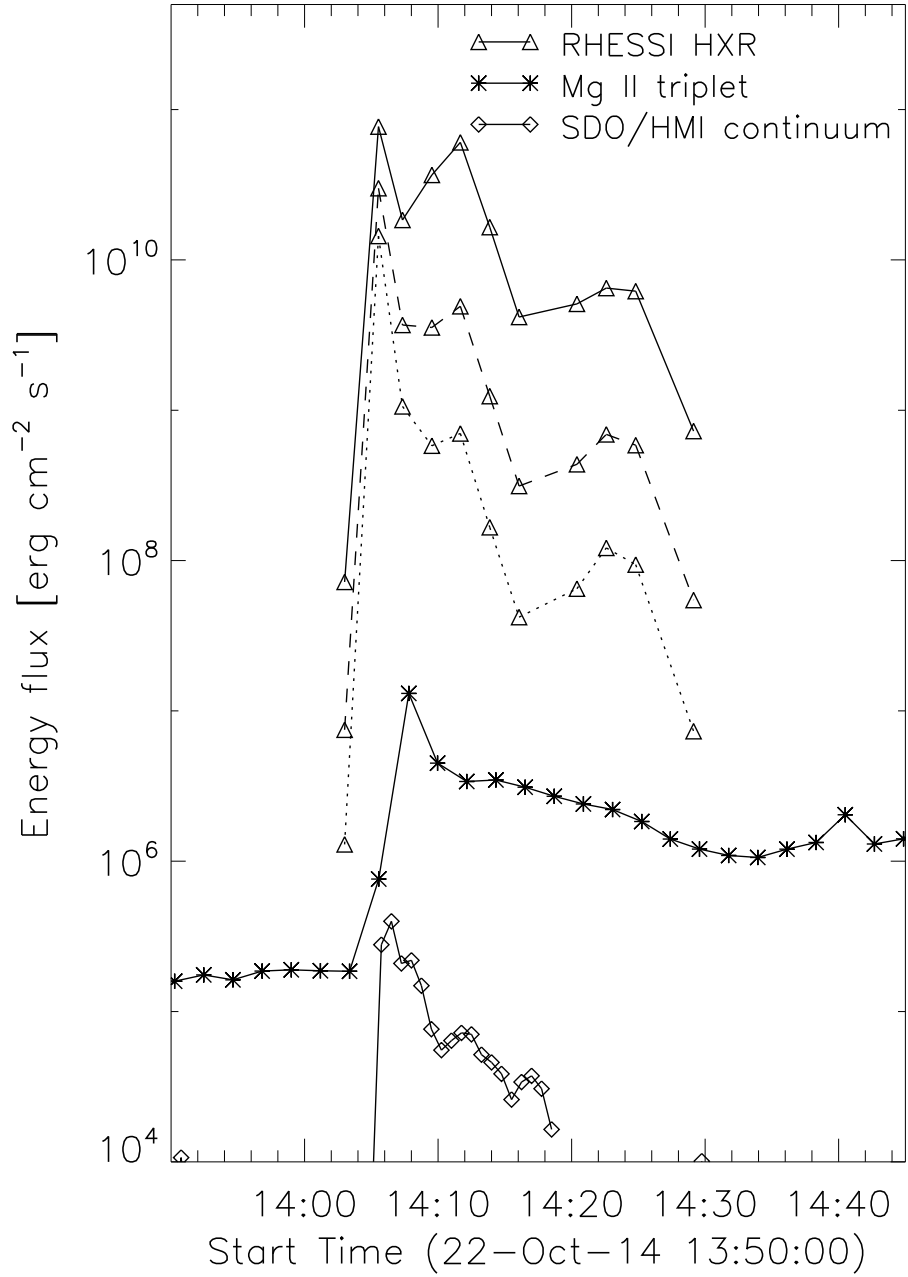} \caption{The energy flux of the bright kernel during the flare impulsive phase estimated from the RHESSI HXR emission with a different threshold energy of 30 keV (solid line), 40 keV (dashed line), and 50 keV (dotted line), the \ion{Mg}{2} triplet intensity observed by IRIS, and the WL continuum emission from SDO/HMI.   \label{fig15}}
\end{figure*}


\end{document}